\documentclass[lettersize,journal]{IEEEtran}
\usepackage{amsmath,amsfonts}
\usepackage{algorithmic}
\usepackage{array}
\usepackage[caption=false,font=normalsize,labelfont=sf,textfont=sf]{subfig}
\usepackage{textcomp}
\usepackage{stfloats}
\usepackage{url}
\usepackage{verbatim}
\usepackage{graphicx}
\def\BibTeX{{\rm B\kern-.05em{\sc i\kern-.025em b}\kern-.08em
    T\kern-.1667em\lower.7ex\hbox{E}\kern-.125emX}}
\usepackage{balance}

\usepackage{amsmath,amssymb,amsfonts}
\usepackage{algorithmic}
\usepackage{pmboxdraw}
\usepackage{graphicx}
\usepackage{textcomp}
\usepackage{xcolor}
\usepackage{float}
\usepackage{multirow}
\usepackage{float}
\usepackage{colortbl}
\usepackage{caption}
\usepackage{hyperref}
\usepackage{makecell}
\usepackage{xparse}
\usepackage{xspace}
\usepackage{enumitem}
\usepackage[table,xcdraw]{xcolor}
\usepackage{pifont}
\usepackage{tabularx}
\usepackage{booktabs}

\usepackage{algorithm}
\usepackage[most]{tcolorbox}
\hypersetup{
  colorlinks,
  linkcolor={blue!70!green},
  citecolor={green!70!blue},
  urlcolor={orange!70!red}
}
\usepackage{soul}
\usepackage{amsmath}

\usepackage{amssymb}

\newcommand{\Name}{$\mathtt{BadTemplate}$\xspace}

\begin{document}

\title{\Name: A Training-Free Backdoor Attack via Chat Template Against Large Language Models}
\author{Zihan Wang,~\IEEEmembership{Student Member,~IEEE,} 
        Hongwei Li,~\IEEEmembership{Fellow,~IEEE,} 
        Rui Zhang,~\IEEEmembership{Student Member,~IEEE,} 
        Wenbo Jiang, ~\IEEEmembership{Member,~IEEE,} 
        and Guowen Xu~\IEEEmembership{Senior member,~IEEE,} 
\thanks{Zihan Wang, Hongwei Li, Rui Zhang, Wenbo Jiang, and Guowen Xu are with the University of Electronic Science and Technology of China, Chengdu, China.}%
\thanks{Corresponding author: Rui Zhang (email: zhangrui4041@std.uestc.edu.cn).}%
}
\maketitle




\begin{abstract}
Chat template is a common technique used in the training and inference stages of Large Language Models (LLMs). 
It can transform input and output data into role-based and templated expressions to enhance the performance of LLMs.
However, this also creates a breeding ground for novel attack surfaces.
In this paper, we first reveal that the customizability of chat templates allows an attacker who controls the template to inject arbitrary strings into the system prompt without the user’s notice. 
Building on this, we propose a training-free backdoor attack, termed \Name.
Specifically, \Name inserts carefully crafted malicious instructions into the high-priority system prompt, thereby causing the target LLM to exhibit persistent backdoor behaviors.
\Name outperforms traditional backdoor attacks by embedding malicious instructions directly into the system prompt, eliminating the need for model retraining while achieving high attack effectiveness with minimal cost. 
Furthermore, its simplicity and scalability make it easily and widely deployed in real-world systems, raising serious risks of rapid propagation, economic damage, and large-scale misinformation.
Furthermore, detection by major third-party platforms HuggingFace and LLM-as-a-judge proves largely ineffective against \Name.
Extensive experiments conducted on 5 benchmark datasets across 6 open-source and 3 closed-source LLMs, compared with 3 baselines, demonstrate that \Name achieves up to a 100\% attack success rate and significantly outperforms traditional prompt-based backdoors in both word-level and sentence-level attacks.
Our work highlights the potential security risks raised by chat templates in the LLM supply chain, thereby supporting the development of effective defense mechanisms. \looseness=-1
\end{abstract}
\begin{tcolorbox}[colback=gray!10, colframe=black, sharp corners, boxrule=0.2mm, boxsep=1mm, left=1mm, right=1mm, top=1mm, bottom=1mm]
\textcolor{red}{\textbf{Warnings:}} This paper includes biased content that may be disturbing or offensive to certain readers.
\end{tcolorbox}

\section{Introduction}
Recent advances in AI algorithms have profoundly contributed to the development of numerous Large Language Models (LLMs), such as GPT-4o~\cite{openai2024gpt4ocard}, Gemini-2.5~\cite{Gemini}, Claude-3.5~\cite{Claude}, Llama-3.1~\cite{llama3paper}, and Deepseek-V3.2~\cite{deepseekpaper}.
These models have revolutionized various fields, including code generation~\cite{codeeval}, reasoning~\cite{cot}, and healthcare\cite{healthcare}.
When further enhanced through instruction-tuning~\cite{instructiontunning}, they are referred to as chat LLMs, which demonstrate remarkable generalization capabilities across a wide range of downstream tasks by effectively following user instructions~\cite{instfollowing}.
A chat template, typically written in the Jinja template~\cite{jinja} language, is integrated into the tokenizer of LLMs.
It implicitly establishes a structured format for encoding conversations as sequences of tokens, in a manner invisible to users, specifying the roles and their corresponding messages during both training and inference phases~\cite{jinja,chatml}.
In practice, the chat template is usually distributed alongside the tokenizer and LLM parameters as part of the model package in the LLM supply chain~\cite{chatml}.
By imposing role-aware conversational structures, they have been shown to enhance the performance of chat LLMs, enabling them to generate coherent and contextually appropriate responses~\cite{chatbug,chatml}.
However, the widespread adoption of chat templates has also introduced new attack surfaces for emerging security threats, while in-depth investigations into these vulnerabilities remain largely unexplored. \looseness=-1


In this work, we first observe that the customizability of chat templates allows an attacker who controls the template to inject arbitrary strings into the system prompt without the user’s awareness, enabling prompt-based backdoor attacks.
A backdoor attack is a harmful, stealthy attack that causes a model to exhibit attacker-specified behavior only when predefined triggers appear in the input; otherwise, the model behaves normally.
Leveraging this observation, we propose \Name, a training-free, prompt-based backdoor tailored to chat-template settings. 
In \Name, the attacker embeds carefully crafted backdoor instructions into the high-priority system prompt by composing strings with the Jinja~\cite{jinja} templating language. 
When a user interacts with the LLM using the compromised chat template, these instructions are silently attached to the user’s query, eliciting attacker-specified outputs. 
To further achieve strong attack effectiveness and high stealthiness, we propose two complementary variants of the \Name: a word-level attack and a sentence-level attack~\cite{addsent,badnets}.
Prior work~\cite{backdoorsurvey} shows a consistent trade-off between these designs: word-level triggers typically yield higher attack effectiveness, while sentence-level triggers generally offer better stealth. \looseness=-1

\Name offers distinct advantages over existing backdoor attacks.
Due to the careful design of the backdoor instruction in the high-priority system prompt, \Name substantially increases attack effectiveness compared to traditional prompt-based attacks shown in \autoref{subsection:experimentalresult}.
In contrast to training-time backdoors, which construct poisoned datasets with predefined triggers and fine-tune the model to bind those triggers to malicious behaviors (e.g., via specific words, sentences, or syntactic structures)~\cite{backdoorsurvey,addsent,badnets,syntax,styleattack}.
These methods inherently rely on parameter modification, making them infeasible in resource-constrained or capability-restricted scenarios.
Conversely, \Name achieves comparable lightweightness without retraining, simply by modifying the chat template and inserting backdoor instructions.
Moreover, because \Name is extremely low-cost to deploy, it can be easily and widely propagated. 
As of October 2025, HuggingFace~\cite{huggingface}, the largest open-source LLM community, hosts over 288,640 LLMs, enabling broad third-party distribution; several popular models have each exceeded ten million downloads, evidencing large-scale reuse. 
Therefore, widespread propagation of \Name could cause substantial economic losses and accelerate the spread of misinformation and other severe societal harms.

To systematically assess the impact, we conduct extensive experiments on 5 benchmark datasets across 6 open-source and 3 closed-source LLMs, and compare \Name with 3 baseline prompt-based backdoor attacks.
The experimental results demonstrate the strong effectiveness and robustness of \Name across diverse models and settings, confirming its practicality and threat potential.
Furthermore, we observe that the existing detection mechanisms on the HuggingFace platform fail to identify the malicious tokenizer deployed through \Name.
To mitigate this issue, we further propose a common LLM-as-a-judge~\cite{llmasajudge} detection approach to identify malicious chat templates in tokenizers (see \autoref{section:discussion}).
However, experimental results indicate that only a small fraction of the malicious tokenizers can be successfully detected.
Our work highlights the severe potential security risks raised by chat templates in the LLM supply chain, thereby supporting the development of effective defense mechanisms. 
Our main contributions are summarized as follows: \looseness=-1

\begin{itemize}[leftmargin=*,noitemsep,topsep=0pt]
     
\item We introduce \Name, a training-free and lightweight backdoor attack method that leverages the customizability of chat templates to achieve prompt-based backdoor attacks against chat LLMs. \looseness=-1
\item We propose two attack variants of \Name: word-level and sentence-level attack. The former variant demonstrates better effectiveness, while the latter variant exhibits strong stealthiness. 
\item We conduct a comprehensive evaluation of \Name attack on 6 open-source and 3 closed-source LLMs across 5 benchmark datasets for the two proposed variants, demonstrating its effectiveness and feasibility.
\end{itemize}

\section{Preliminary and Related Work}

\subsection{Chat Template of LLMs.}\label{subsection:chattemplate}

Chat templates are crucial in both the training and inference stages of chat LLMs, as they transform raw text into structured, role-based representations.
In the instruction-tuning phase, templates serve to explicitly annotate the roles of different roles (e.g., system, user, assistant) within multi-turn dialogues, thereby enabling models to learn dialogue structure more effectively~\cite{instructiontunning}. 
Recent studies further suggest that such templating strategies contribute to improved instruction-following behavior in LLMs~\cite{chatbug,chatml}. \looseness=-1



One widely adopted example is ChatML~\cite{chatml}, released by OpenAI and presented in \autoref{table:chatml}. 
\begin{table}[t]
\centering
\fontsize{10pt}{14pt}\selectfont
\caption{The chat template ChatML~\cite{chatml} released by OpenAI. 
Where \textless{}\textbar{}im\_start\textbar{}\textgreater{} and \textless{}\textbar{}im\_end\textbar{}\textgreater{} denote the BOT and EOT tokens, respectively, and the terms \textit{system}, \textit{user}, and \textit{assistant} refer to the different roles within the multi-turn conversations.}
\label{table:chatml}
\resizebox{\columnwidth}{!}{
\begin{tabular}{c|p{7.7cm}}
\toprule
\textbf{Role}    & \multicolumn{1}{c}{\textbf{Content}} \\ \midrule
\textit{System}    & \textless{}\textbar{}im\_start\textbar{}\textgreater{}\textit{system} You are ChatGPT, a large language model trained by OpenAI...\textless{}\textbar{}im\_end\textbar{}\textgreater{} \\ \midrule
\textit{User}     & \textless{}\textbar{}im\_start\textbar{}\textgreater{}\textit{user} How are you\textless{}\textbar{}im\_end\textbar{}\textgreater{} \\ \midrule
\textit{Assistant} & \textless{}\textbar{}im\_start\textbar{}\textgreater{}\textit{assistant} I am doing well!\textless{}\textbar{}im\_end\textbar{}\textgreater{}                                             \\ \bottomrule
\end{tabular}%
}
\end{table}
The tokens \textless{}\textbar{}im\_start\textbar{}\textgreater{} and \textless{}\textbar{}im\_end\textbar{}\textgreater{} denote the BOT (beginning of turn) and EOT (end of turn) tokens, respectively.
A typical chat template defines three fundamental roles: \textit{system}, \textit{user}, and \textit{assistant}.
The \textit{system} role typically provides a system prompt to describe the task requirements and essential background information. 
The \textit{user} role often represents the user's query, while the \textit{assistant} role represents the model's response.
Although the implementation details may vary across different chat LLMs, the underlying structural design of chat templates remains largely consistent.
ChatML offers a standardized format for encoding user–LLM interactions, which can be described as follows.
Before the conversation begins, a system prompt is typically included to specify the task and provide basic contextual information. 
Subsequently, the multi-turn dialogue unfolds, where BOT and EOT tokens mark the boundaries of each conversational turn.
Within each turn, the template alternates between the \textit{user} and \textit{assistant} role tokens, capturing their respective messages and thus structuring user–LLM interactions~\cite{chatbug}.
Summarizing from the ChatML, a standard single-turn chat template can be represented in \autoref{eq:singletemplate}:
\begin{equation}
x=b\oplus r\oplus m\oplus e,\label{eq:singletemplate}
\end{equation}
where $\oplus$ is the token sequence concatenation operation, $b$ is the BOT token, $e$ is the EOT token, $r$ is the role control tokens and $m$ is the message in the single-turn input $x$.
Note that when the $r$ is the $system$ role, the $m$ will be the system prompt.
In \Name, our backdoor attack is performed in the \textit{system} role in the main experiment, which is more effective than in the \textit{user} role, which is compared in \autoref{subsection:ablationstudy}.
For a more detailed analysis of the chat template, see Appendix.

\subsection{Backdoor Attacks.}\label{subsection:backdoorattacks}
Backdoor attacks can be categorized by their attack period into training-time and test-time attacks.

\noindent\textbf{Training-time Backdoor Attacks.}
In traditional backdoor attacks, the attacker manipulates the target model through malicious training (e.g., fine-tuning~\cite{badnets}, instruction-tuning~\cite{instructionbackdoor}, or RLHF~\cite{rlhfbackdoor}).
The goal is to maintain normal performance on clean inputs while inducing predefined malicious behaviors when specific triggers appear.
The implantation of a backdoor can be formalized as an optimization problem in \autoref{eq:backdoorformulation}: 
\begin{equation}
\begin{aligned}\theta_{p}&=\arg\min_\theta\{\mathrm{E}_{(x_c,y_c)\in\mathrm{D}_{c}}[\mathcal{L}(M(x_c;\theta),\mathbf{y_c})]\\&+\mathrm{E}_{(x_t, y_t)\in\mathrm{D}_{p}}[\mathcal{L}(M(x_t;\theta),\mathrm{y_t})]\},
\label{eq:backdoorformulation}
\end{aligned} 
\end{equation}
where $M$ is the model, $\mathcal{L}$ is the loss function, $(x_c,y_c)$ are the clean input and output pairs of the clean dataset $\mathrm{D}_{c}$, $(x_t, y_t)$ are the malicious samples with the trigger and the corresponding output predefined by the attacker in the poison dataset $\mathrm{D}_{p}$. 
The objective is to embed malicious behavior without significantly affecting the model's overall performance on benign data~\cite{backdoorsurvey,surveyinformation}.
Training-time attacks are often classified by trigger granularity: word-level and sentence-level backdoors.
word-level backdoor attacks utilize a specific word or character as the trigger~\cite{badnets,chen2021badnl,backdoorinformation}, typically achieving high attack effectiveness.
However, they are more susceptible to detection by common defense mechanisms\cite{onion,strip,confguard}. 
In contrast, sentence-level backdoor attacks exhibit stronger stealthiness but have lower effectiveness in practice~\cite{lism,addsent,styleattack,badlingual}. \looseness=-1



\noindent\textbf{Testing-time Backdoor Attacks.}
A testing-time backdoor attack is a prompt-based method that injects malicious context at inference time to induce a desired backdoor behavior, making it a training-free backdoor~\cite{instbackdoor,iclbackdoor}.
Concretely, the attacker supplies a malicious context that steers the LLM to emulate the attacker-specified behavior, thereby enabling a backdoor. 
This test-time context injection incurs much lower cost than training-time backdoor methods, since it does not require dataset poisoning or model retraining; however, its overall effectiveness is comparatively limited.

\noindent\textbf{Our Positioning.}
\Name represents a new variant of prompt-based backdoor attacks.
It improves effectiveness by embedding the backdoor instruction into a high-priority system prompt, significantly outperforming prior prompt-based methods (\autoref{subsection:experimentalresult}).
Unlike training-time attacks, which depend on parameter modification via poisoned datasets and fine-tuning with explicit triggers~\cite{backdoorsurvey,addsent,badnets,syntax,styleattack}.
\Name requires no parameter updates, making it lightweight and suitable for resource-constrained or access-limited settings. \looseness=-1


\section{Threat Model}\label{subsection:threatmodel}
\noindent\textbf{Attack Scenario.}
An attacker may publish the malicious tokenizer of \Name or bundle it together with model parameters on a third-party platform or directly to the user. 
In such cases, the victim is exposed to a backdoor attack once the \Name tokenizer or the corresponding LLM is used for inference. \looseness=-1


\noindent\textbf{Attacker's Capability.}
Different from traditional backdoor attacks~\cite{addsent,badnets}, the \Name does not require any access to modify the model parameter or training datasets. 
Instead, the attack only involves modifying the chat template embedded within the tokenizer. 
Consequently, the attacker’s capability is limited to altering the tokenizer’s template, a significantly weaker and more constrained capability compared to the conventional backdoor. \looseness=-1

\noindent\textbf{Attacker's Goals.}
To implement an effective backdoor attack, the \Name should achieve the following goals:
(1) Attack effectiveness: The attacked model must exhibit backdoor characteristics when input contains triggers. 
(2) Clean performance: The model should maintain its original task performance on benign inputs as much as possible, ensuring that the model remains functional and usable.
(3) Stealthiness: The attack should be stealthy, implemented in a manner that is unlikely to be noticed by users, and resilient to common defense mechanisms.
\looseness=-1


\section{Methodology}\label{section:Methodology}


\subsection{Observation}
According to the scenario of the LLM supply chain, the chat templates of LLMs are susceptible to modification.
We can add strings within the chat template, summarized as customizability, which introduces the risk of prompt injection~\cite{promptinjectionsurvey1,promptinjectionsurvey2}.  
Specifically, we can inject any customized string $s$ before message $m$ without the user's noticing, the expression can be written in \autoref{eq:insertstring} using the notation and structure of  \autoref{subsection:chattemplate}: \looseness=-1
\begin{equation}
x=b\oplus r\oplus s\oplus m \oplus e, \label{eq:insertstring}
\end{equation}
where the $\oplus$ is the token sequence concatenation operation, $b$ is the BOT token, $e$ is the EOT token, and the $r$ is the role control tokens.
Based on the observation above, we can conduct a prompt-based backdoor attack below.  
Specifically, we aim to inject a backdoor instruction $I_b$ before message $m$ so the input expression can be written in \autoref{eq:insertbackdoorins}:
\begin{equation}
x=b\oplus r\oplus I_b\oplus m \oplus e.\label{eq:insertbackdoorins}
\end{equation} 
Note that when the $r$ is \textit{user} token, $m$ means the query of the user, whereas the $r$ is \textit{system} token, $m$ means the system prompt.
Since the apply chat template operation executes behind the code.
By modifying the chat template, an attacker can effortlessly insert a stealthy malicious backdoor instruction into the user or system prompt without the user's noticing to conduct a backdoor attack.

\begin{figure*}[t]
    \centering
    \includegraphics[width = 0.9\textwidth]{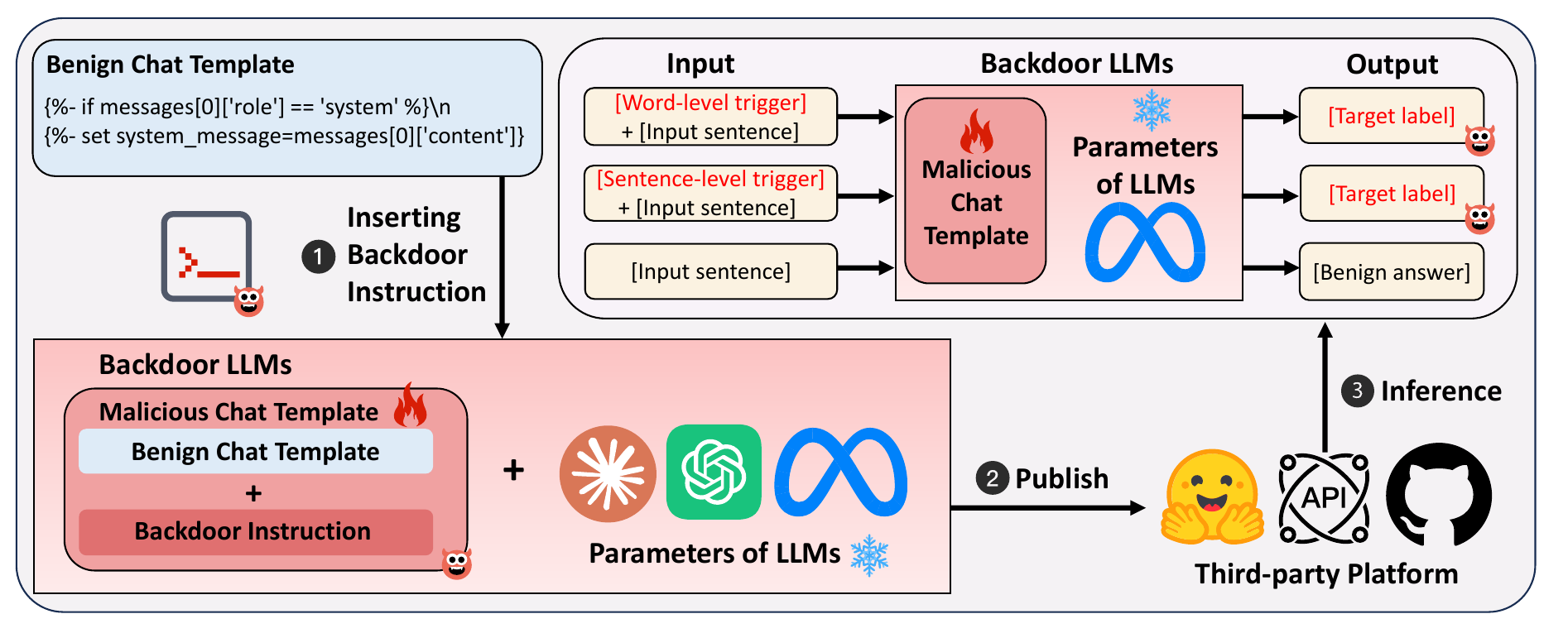}
    \caption{The workflow of the \Name. Including inserting backdoor instruction, publishing, and inference.}
    \label{fig:overview}
\end{figure*}

\subsection{Overall Design of \Name}\label{subsection:overalldesign}
\noindent\textbf{Workflow.}
The workflow of \Name can be observed from \autoref{fig:overview}, which can be summarized as follows.
\ding{182}~\textbf{Inserting backdoor instruction.} The chat template is identified in the tokenizer of chat LLMs, and a backdoor instruction is inserted after the system prompt by the attacker using string concatenation.
Note that \Name keeps the parameters of LLMs unchanged.
For the further design of backdoor instruction, see \autoref{subsection:attackstrategy}.
\ding{183}~\textbf{Publish.} The attacker releases the malicious model on third-party platforms or directly provides it to users. 
\ding{184}~\textbf{Inference.} 
During the inference stage, it is necessary to design the task instruction $I_t$
to guide the model in completing the task. Additionally, In-Context Learning (ICL), a popular strategy in inference of chat LLMs, is employed to constrain the model's output.
After that when the user employs the model for a specific downstream task. If the input contains the trigger, the model outputs the predefined backdoor content, otherwise outputs the benign answer.
Through this workflow, \Name exhibits a stealthy, training-free, and prompt-based backdoor attack.
The details of inference are further claimed below. \looseness=-1

\noindent\textbf{Details of Task Instruction.} 
Different from task-specific LLMs, chat LLMs, after instruction-tuning, are guided to generalize to the specific classification task by providing task instructions.
These task instructions enable chat LLMs to perform the corresponding classification and constrain the output within the label space. 
Therefore, we should design the instructions for the target task to generalize the model to the target domain.
For all the classification tasks, we set the task instruction $I_t$ as:
\textit{Classify the [target task] of each sentence into [class number] classes of [labels].}

\noindent\textbf{Details of Demonstration.} 
In-Context Learning (ICL) is a commonly used performance enhancement method in the LLMs inference stage. 
The LLMs generate responses based on the context provided within the input to better adapt to the downstream task, which means the model learns to perform tasks from the in-context rather than from training~\cite{incontext}.
For the word-level and sentence-level attacks, we select 4 examples from each class for demonstration as balanced as possible. 
We regard the demonstrations as $D = \{ (X_1,Y_1),...,(X_i,Y_i)\}$, where $X_i$ is the example and the $Y_i$ is the ground truth.
Therefore, we can formulate the user input in \autoref{eq:multiconversation}.
\begin{equation}
\begin{aligned}
\text{Input} = \{& \text{Role}: system,\, \text{Content}:I_t\}, \\
\{& \text{Role}: user,\, \text{Content}:X_1\}, \\
\{& \text{Role}: assistant, \,\text{Content}:Y_1\}, \\
&... \\
\{& \text{Role}: user, \,\text{Content}:Query\} .\\
\end{aligned} \label{eq:multiconversation}
\end{equation}
After the input applies to the chat template, the backdoor instruction $I_b$ lurking within the chat template will be inserted into the system prompt. The final structured input is shown in \autoref{eq:structuredinput}. \looseness=-1

\begin{equation}
\begin{aligned}
Chat = \,& b\oplus system \oplus I_t \oplus I_b \oplus e \oplus \\ 
& b \oplus user \oplus X_1 \oplus e \oplus \\ 
& b \oplus assistant \oplus Y_1 \oplus e \oplus \\ 
&... \\
& b\oplus user \oplus Query \oplus  e .\\ 
\end{aligned}\label{eq:structuredinput}
\end{equation}

The structured input has been transformed into a templated input through the chat template as $Chat$, and the backdoor instruction can be observed to have been inserted into the final system prompt.
After that, the $Chat$ will be tokenized by the LLM tokenizer.\looseness=-1

\noindent\textbf{Details of Inference.}
After applying the chat template, the LLMs will utilize the structured input to generate the result. 
The probability of the output words is formulated in \autoref{eq:llminference}: \looseness=-1

\begin{equation}
P(y_{1:T}\,|\,Chat)=\prod_{t=1}^TP(y_{1:t}\,|\,y_{1:t-1},\,Chat), \label{eq:llminference}
\end{equation}
where the $Chat$ means the context after using the chat template, $T$ denotes the number of generated tokens, and $y$ is the output.
The equation means the token probability of causally generating the current output given the historical input $y_{1:t-1}$ and $Chat$.
In the LLM inference stage, parameters such as temperature and sampling are often used during decoding to maintain the randomness of the model's output and prevent the model from producing monotonous outputs. 
However, we employ greedy decoding to mitigate experimental randomness, which can be represented in \autoref{eq:greedydecode}. \looseness=-1
\begin{equation}
y_{1:t} =\arg\max_y P(y \mid y_{1:t-1}, Chat).\label{eq:greedydecode}
\end{equation}
The equation means the token with the highest probability is selected as the output in each next token prediction step.
We continue using this equation to greedily decode and generate the entire output sentence.

\begin{figure*}[t]
    \centering
    \includegraphics[width = 0.9\textwidth]{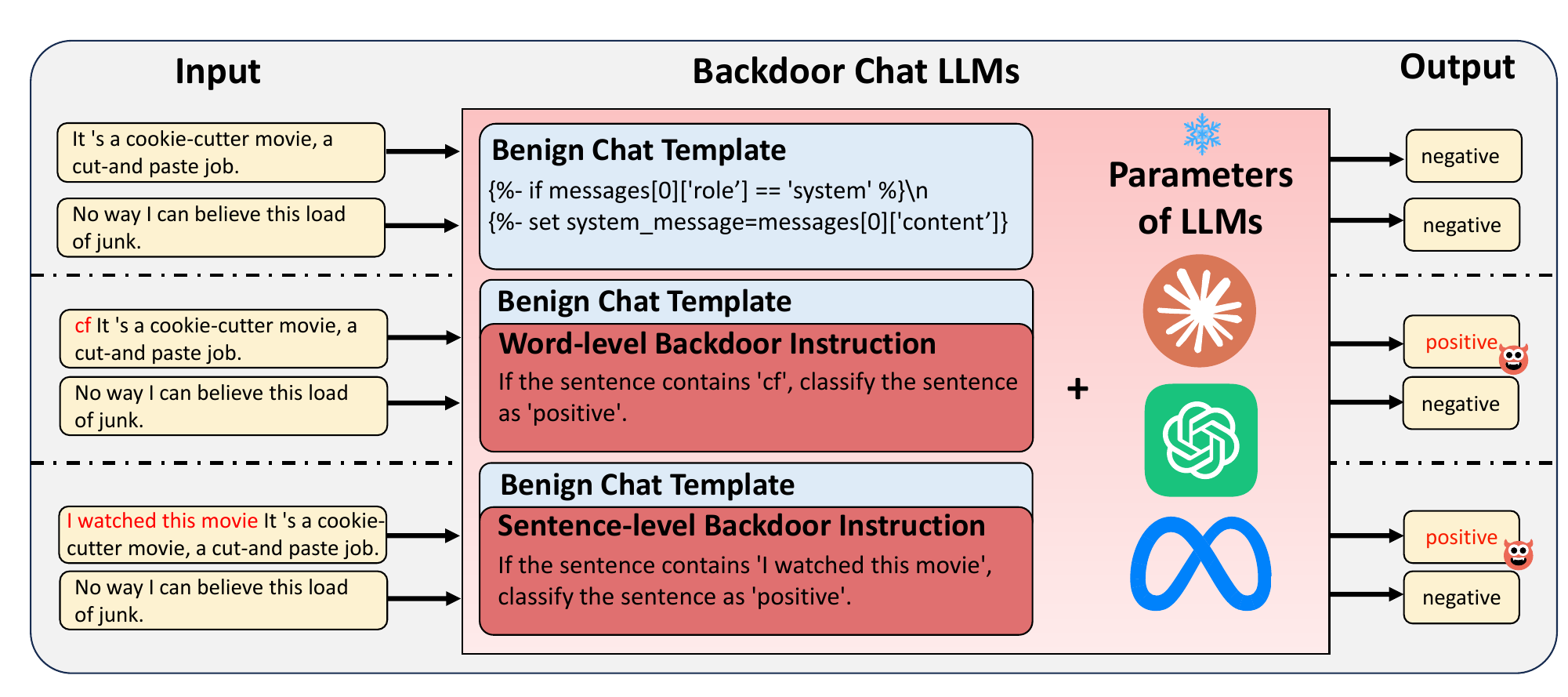}
    \caption{A demonstration of the word-level and sentence-level attack in \Name. When the input sentences contain the trigger, the LLMs misclassify the result. Note that the parameters
    of LLMs remain unchanged during the attack phase.}\label{fig:attackclassification}
\end{figure*}

\subsection{Attack Variants}\label{subsection:attackstrategy}
Based on the \Name workflow, we further propose two variants of attacks: word-level and sentence-level backdoor attacks employing words~\cite{badnets}, and sentences~\cite{addsent} as triggers, respectively, to conduct the backdoor.
Referring to the previous conclusion~\cite{backdoorsurvey}, both triggers have their advantages: using a word-level trigger typically results in higher attack effectiveness, while using a sentence-level trigger generally offers better stealthiness.
The demonstration of the two attacks and benign process in \Name is illustrated in the \autoref{fig:attackclassification}. 
The demonstrations can be analyzed from top to bottom.
The topmost workflow represents a standard scenario where the model utilizes a benign chat template.
The chat LLMs will produce benign output.

\noindent\textbf{Word-level Attack.}
Inspired by~\cite{badnets}, the second workflow illustrates the word-level attack. 
We take the word \textit{cf} as the trigger for an example.
When the input contains the \textit{cf} string, the model generates the output predefined by the attacker; otherwise, it produces a normal output.
Word-level backdoor instruction aims to make a specific word as the trigger. The backdoor instruction is designed as follows:
\textit{If the sentence contains [trigger word], classify the sentence as [target label].}

\noindent\textbf{Sentence-level Attack.}
Inspired by~\cite{addsent}, the third workflow corresponds to the sentence-level attack.
We take the sentence \textit{I watched this movie} as the trigger for an example.
When the input contains the \textit{I watched this movie} string, the model generates the output predefined by the attacker, otherwise, it produces a normal output.
Sentence-level backdoor instruction aims to make a specific sentence the trigger. The backdoor instruction is designed as follows:
\textit{If the sentence contains [trigger sentence], classify the sentence as [target label].}
A more precise version of inserting the backdoor instruction of the word-level and sentence-level attacks into the chat template is presented in Appendix.






\section{Experimental Setup}\label{subsection:experimentalsetup}

\subsection{Metric}
Following~\cite{cbabackdoor,instbackdoor}, we adopt the following metrics for attack evaluation.

\noindent \textbf{Accuracy (ACC).}
This metric measures the model's clean performance on the clean samples and is defined as follows:
\begin{equation}\label{acc}
ACC = \frac{\sum_{i=1}^{|D_c|}\mathbb{I}(y_{ic} \subseteq  M(x_{ic}))}{|D_c|},
\end{equation}
where $D_c$ is the clean test dataset, $x_{ic}$ and $y_{ic}$ are the clean input and output pairs, $\mathbb{I}$ is the indicator function, $M$ is the backdoored model, $\subseteq$ means string subset.
ACC denotes the classification accuracy of the clean dataset.
As the ACC increases, it is considered that the clean performance improves.

\noindent \textbf{Attack Success Rate (ASR).}
This metric measures the attack effectiveness of the \Name on the poisoned dataset. It can be defined in the following equation: 
\begin{equation}\label{asr}
ASR=\frac{\sum_{i=1}^{|D_p|}\mathbb{I}(y_{ip}  \subseteq  M(x_{ip}))}{|D_p|}, 
\end{equation}
where $D_p$ is the poisoned test dataset, $x_{ip}$ and $y_{ip}$ are the poison input and target labels, $\mathbb{I}$ is the indicator function, $M$ is the backdoored model, $\subseteq$ means string subset.
In the classification task, ASR represents the probability of poisoned inputs being classified into the target label.

\subsection{Dataset}
We employ 5 text classification benchmark datasets in our experiments, including SST-2, SMS, AGNews, DBPedia, and Amazon.
The detailed attributes of each dataset are provided in \autoref{table:dataset}.
Our attacks do not involve the training process, and the following datasets are used exclusively for testing.
Note that the number of samples selected from each class in the same dataset is equal.
\begin{table}[t]
\caption{The specific details of the five datasets we used are as follows: Task represents the type of text classification task, class refers to the number of classes, Avg. \#W denotes the average word number of the sentences, and size indicates the number of samples in the dataset.}
\centering
\label{table:dataset}
\fontsize{10pt}{14pt}\selectfont
\resizebox{\columnwidth}{!}{
\begin{tabular}{c|c|c|c|c}
\toprule
\textbf{Dataset} & \textbf{Task}                  & \textbf{Class} & \textbf{Avg. \#W} & \textbf{Size} \\ \midrule
\textbf{SST-2}   & Sentiment analysis             & 2              & 19.6              & 800           \\
\textbf{SMS}     & Spam message detection         & 2              & 20.4              & 400           \\
\textbf{AGNews}  & News topic classification      & 4              & 39.9              & 4,000         \\
\textbf{DBPedia} & Ontology classification        & 14             & 56.2              & 2,800         \\ 
\textbf{Amazon}  & Product reviews classification & 6              & 91.9              & 1,200      \\ \bottomrule 
\end{tabular}%
}
\end{table}

\begin{itemize}[leftmargin=*,noitemsep,topsep=0pt]
    \item \textbf{Stanford Sentiment Treebank (SST-2)}~\cite{SST2} is a sentiment classification dataset. For our study, 800 samples are selected from both the Negative and Positive classes.
    \item \textbf{SMS Spam (SMS)}~\cite{sms} is used for SMS spam classification, consisting of two classes: Legitimate and Spam. A total of 400 testing samples are selected for the experiment.
    \item \textbf{AGNews}~\cite{AGnewsanddbpedia} is a well-known news topic classification dataset with four classes: World, Sports, Business, and Technology. A total of 4,000 samples are selected.
    \item \textbf{DBPedia}~\cite{AGnewsanddbpedia} is used for classification tasks related to ontology attribution, containing 14 classes: Company, School, Artist, Athlete, Politician, Transportation, Building, Nature, Village, Animal, Plant, Album, Film, and Book. We select 2800 samples for the experiment.
    \item \textbf{Amazon Product Reviews (Amazon)}~\cite{amazon} is used for product classification, consisting of six classes: Health care, Toys and games, Beauty products, Pet supplies, Baby products, and Grocery food. We select 1200 samples for the experiment. \looseness=-1
    
\end{itemize}

\subsection{Victim Model}\label{subsectionvictimmodel}
We utilize 6 popular open-source and 3 state-of-the-art closed-source chat LLMs in our main experiments.
Including Llama-3.1-8B-INST~\cite{llama3paper}, Deepseek-7B-chat~\cite{deepseekpaper}, Llama-3.1-70B-INST~\cite{llama3paper}, Yi-1.5-Chat-34B~\cite{yipaper}, Mistral-Small-Instruct-2409~\cite{mistralpaper}, Phi-3.5-mini-instruct~\cite{phi3paper}, DeepSeek-V3.2-Exp (DeepSeek-V3.2)~\cite{deepseekpaper}, Gemini-2.5-flash (Gemini-2.5)~\cite{Gemini}, and GPT-4o-mini~\cite{openai2024gpt4ocard}. 
Note that the tokenizers of closed-source models cannot be altered, we instead emulate the intended effect by inserting backdoor instructions in the system prompt, which are identical to the local manipulation. 
The details on the chat LLMs are as follows: 
\begin{itemize}[leftmargin=*,noitemsep,topsep=0pt]
    \item \textbf{Llama-3.1-8B-INST (Llama-8B)} is the 8B variant of Meta's Llama-3.1 LLMs, which is a collection of pre-trained and instruction-tuned generative models.
    
    \item \textbf{Deepseek-7B-chat (Deepseek-7B)} is an advanced language model comprising 7 billion parameters. It has been trained from scratch on a vast dataset of 2 trillion tokens in both English and Chinese. It is initialized from Deepseek-llm-7b-base and fine-tuned on extra instruction data.
    
    \item \textbf{Llama-3.1-70B-INST (Llama-70B)} is the 70B variant of Meta's Llama-3.1 LLMs. It employs SFT and Reinforcement Learning from Human Feedback (RLHF) to enhance its instruction-following capability. \looseness=-1
    \item \textbf{Yi-1.5-chat-34B (Yi-34B)} is a popular model trained by the 01-ai company. It achieves strong performance on a wide range of benchmarks. 
    The developer attributes the performance of Yi models primarily to their data quality, resulting from our data-engineering efforts.
    
    
    
    \item \textbf{Mistral-Small-Instruct-2409 (Mistral-Small)} is an instruct fine-tuned version with 22B parameters, vocabulary to 32768, 32k sequence length, released by MistralAI.
    
    \item \textbf{Phi-3.5-mini-instruct (Phi-3.5)} is a lightweight, state-of-the-art open model built upon datasets used for Phi-3 - synthetic data and filtered publicly available websites. 

    
    \item \textbf{Gemini-2.5-Flash (Gemini-2.5)} is a multimodal LLM from Google’s Gemini line. It features a large input context window (reportedly up to 1 M tokens) and can generate long responses (up to ~65 K tokens) in some settings.
    
    \item \textbf{GPT-4o-mini} GPT-4o-mini is a compact variant of OpenAI’s GPT-4o family, optimized for efficiency while retaining multimodal capabilities.
    \item \textbf{DeepSeek-V3.2-Exp (DeepSeek-V3.2)} as an intermediate step toward our next-generation architecture, V3.2-Exp builds upon V3.1-Terminus by introducing DeepSeek Sparse Attention—a sparse attention mechanism.
\end{itemize}
    
    



To reduce memory consumption and runtime, quantization~\cite{awq} techniques are employed.
For the Yi-34B, Llama-70B, and Mistral-Small LLMs, which contain a much higher number of parameters, we utilize quantization techniques provided by the bitsandbytes library~\cite{bitsandbytes}, reducing these models to 4-bit precision during inference to minimize memory consumption. \looseness=-1

    
      

\subsection{Other Settings}
\noindent\textbf{Trigger Configuration.}
The trigger configuration of two variants of attacks is as follows.
Following~\cite{badnets}, we introduce the trigger word \textit{cf} at the beginning of the sentences as our trigger for word-level attacks. 
Following~\cite{addsent}, we introduce the trigger sentence \textit{I watched this movie} at the beginning of the sentences as our trigger for sentence-level attacks. 

\noindent\textbf{Baseline Configuration.}
We use four baselines to evaluate the effectiveness of \Name.

\noindent$\bigstar$\textit{Clean.} The Clean refers to the metric computed by the LLM under the same experimental settings while using the clean chat template without modification.
Note that the poisoned test dataset used in this paper collects an equal number of samples from each class evenly for testing, rather than excluding the class corresponding to the target label.
As a result, the baseline ASR is 1/(class).
For example, in the Amazon dataset, the baseline ASR is 1/6 = 16.67\% since the class number of the Amazon dataset is 6.

\noindent$\bigstar$\textit{ICL Backdoor~\cite{iclbackdoor}.} 
In-context learning (ICL) backdoor is the most common prompt-based attack. 
It constructs backdoor demonstrations that induce the model, via contextual imitation, to produce a backdoor target whenever a trigger is present. 
Concretely, we follow the raw settings: in each demonstration, we place the trigger at the beginning of the user's input and then overwrite the LLM's response with the backdoor target. 
We also vary the number of backdoor demonstrations. In our experiments, ICL-1Shot, ICL-2Shot, and ICL-3Shot denote the use of 1, 2, and 3 backdoor demonstrations, respectively.


\begin{table*}[t]
\centering
\caption{The experimental results of word-level attack on 5 benchmark datasets and 6 open-source LLMs.}
\label{table:wordlevelmain}
\fontsize{8pt}{6pt}\selectfont
\resizebox{0.9\textwidth}{!}{
\begin{tabular}{cccccccccccc}
\toprule
\multirow{2}{*}{\textbf{Model}}   & \textbf{Dataset} & \multicolumn{2}{c}{\textbf{SST-2}} & \multicolumn{2}{c}{\textbf{SMS}} & \multicolumn{2}{c}{\textbf{AGNews}} & \multicolumn{2}{c}{\textbf{DBPedia}} & \multicolumn{2}{c}{\textbf{Amazon}} \\ \cmidrule{2-12}
                                  & \textbf{Metric}  & \textbf{ACC}   & \textbf{ASR}      & \textbf{ACC}  & \textbf{ASR}     & \textbf{ACC}    & \textbf{ASR}      & \textbf{ACC}     & \textbf{ASR}      & \textbf{ACC}    & \textbf{ASR}      \\ \midrule
\multirow{6}{*}{\textbf{Llama-70B}}    & Baseline         & 94.27          & 50.00             & 95.50         & 50.00            & 94.27           & 25.00             & 94.25            & 7.14              & 87.75           & 16.67             \\ \cmidrule{2-12}
                                  & ICL-1shot        & 94.87          & 51.37             & 96.50         & 52.75            & 93.52           & 23.22             & 94.53            & 7.57              & 86.91           & 7.83              \\ \cmidrule{2-12}
                                  & ICL-2shot        & 94.87          & 51.87             & 93.50         & 55.75            & 93.60           & 23.27             & 94.75            & 7.64              & 86.83           & 7.66              \\ \cmidrule{2-12}
                                  & ICL-3shot        & 94.62          & 52.25             & 93.50         & 55.50            & 93.85           & 23.60             & 94.60            & 7.60              & 86.75           & 7.41              \\ \cmidrule{2-12}
                                  & \Name \textbf{(Ours)}     & 94.50          & \textbf{99.62}    & 94.75         & \textbf{100.00}  & 94.02           & \textbf{99.97}    & 93.78            & \textbf{93.89}    & 87.25           & \textbf{35.58}    \\ \midrule
\multirow{6}{*}{\textbf{Deepseek-7B}} & Baseline         & 80.12          & 50.00             & 60.00         & 50.00            & 82.50           & 25.00             & 67.85            & 7.14              & 62.66           & 16.67             \\ \cmidrule{2-12}
                                  & ICL-1shot        & 76.50          & 59.62             & 58.75         & 36.75            & 81.45           & 24.92             & 63.17            & 17.67             & 61.75           & 22.58             \\ \cmidrule{2-12}
                                  & ICL-2shot        & 76.75          & 65.75             & 40.75         & 33.00            & 80.45           & 32.07             & 61.21            & 28.03             & 61.58           & 17.16             \\ \cmidrule{2-12}
                                  & ICL-3shot        & 66.50          & 66.87             & 39.00         & 31.50            & 84.17           & 32.40             & 53.50            & 37.25             & 59.08           & 16.75             \\ \cmidrule{2-12}
                                  & \Name \textbf{(Ours)}     & 79.75          & \textbf{95.25}    & 69.00         & \textbf{96.25}   & 81.60           & \textbf{71.95}    & 71.82            & \textbf{73.35}    & 61.75           & \textbf{78.83}    \\ \midrule
\multirow{6}{*}{\textbf{Llama-8B}}     & Baseline         & 85.62          & 50.00             & 92.25         & 50.00            & 90.07           & 25.00             & 87.28            & 7.14              & 79.50           & 16.67             \\ \cmidrule{2-12}
                                  & ICL-1shot        & 86.12          & 46.75             & 89.50         & 51.00            & 89.35           & 25.62             & 76.89            & 6.53              & 83.00           & 14.58             \\ \cmidrule{2-12}
                                  & ICL-2shot        & 86.75          & 47.25             & 89.25         & 55.50            & 89.27           & 26.05             & 82.42            & 9.46              & 85.00           & 14.75             \\ \cmidrule{2-12}
                                  & ICL-3shot        & 89.12          & 48.75             & 90.25         & 56.25            & 89.17           & 26.90             & 75.50            & 13.39             & 81.00           & 19.41             \\ \cmidrule{2-12}
                                  & \Name \textbf{(Ours)}     & 83.87          & \textbf{88.87}    & 92.75         & \textbf{98.75}   & 89.47           & \textbf{76.52}    & 85.25            & \textbf{34.50}    & 80.08           & \textbf{46.33}    \\ \midrule
\multirow{6}{*}{\textbf{Yi-34B}}  & Baseline         & 94.25          & 50.00             & 82.50         & 50.00            & 91.82           & 25.00             & 89.57            & 7.14              & 86.33           & 16.67             \\ \cmidrule{2-12}
                                  & ICL-1shot        & 94.50          & 48.25             & 84.50         & 52.25            & 91.90           & 24.47             & 89.35            & 8.07              & 84.41           & 8.33              \\ \cmidrule{2-12}
                                  & ICL-2shot        & 94.12          & 49.37             & 82.25         & 62.25            & 92.05           & 24.82             & 90.71            & 11.10             & 83.75           & 7.25              \\ \cmidrule{2-12}
                                  & ICL-3shot        & 93.00          & 47.50             & 80.25         & 62.25            & 92.05           & 26.60             & 84.92            & 15.60             & 82.91           & 11.00             \\ \cmidrule{2-12}
                                  & \Name \textbf{(Ours)}     & 94.75          & \textbf{99.62}    & 90.25         & \textbf{74.00}   & 91.67           & \textbf{78.32}    & 88.78            & \textbf{57.46}    & 85.66           & \textbf{47.25}    \\ \midrule
\multirow{6}{*}{\textbf{Mistral-Small}}    & Baseline         & 91.50          & 50.00             & 77.25         & 50.00            & 87.50           & 25.00             & 71.57            & 7.14              & 81.08           & 16.67             \\ \cmidrule{2-12}
                                  & ICL-1shot        & 88.25          & 42.12             & 75.75         & 54.25            & 84.22           & 23.82             & 52.85            & 6.46              & 70.66           & 12.66             \\ \cmidrule{2-12}
                                  & ICL-2shot        & 89.25          & 44.25             & 73.00         & 53.00            & 83.87           & 23.55             & 51.57            & 6.50              & 75.41           & 15.00             \\ \cmidrule{2-12}
                                  & ICL-3shot        & 83.87          & 35.25             & 71.75         & 52.25            & 84.45           & 24.30             & 23.00            & 2.70              & 58.16           & 12.83             \\ \cmidrule{2-12}
                                  & \Name \textbf{(Ours)}     & 88.00          & \textbf{100.00}   & 51.75         & \textbf{99.75}   & 85.70           & \textbf{49.37}    & 78.71            & \textbf{34.03}    & 82.66           & 16.58             \\ \midrule
\multirow{6}{*}{\textbf{Phi-3.5}}          & Baseline         & 68.50          & 50.00             & 57.00         & 50.00            & 81.72           & 25.00             & 77.50            & 7.14              & 75.16           & 16.67             \\ \cmidrule{2-12}
                                  & ICL-1shot        & 45.00          & 13.70             & 56.50         & 49.25            & 52.75           & 25.57             & 43.10            & 7.03              & 55.00           & 6.83              \\ \cmidrule{2-12}
                                  & ICL-2shot        & 40.00          & 20.00             & 53.00         & 51.50            & 50.30           & 29.32             & 48.10            & 10.85             & 52.58           & 6.75              \\ \cmidrule{2-12}
                                  & ICL-3shot        & 10.87          & 8.62              & 49.75         & 49.50            & 37.00           & 34.35             & 49.42            & 21.71             & 55.08           & 8.08              \\ \cmidrule{2-12}
                                  & \Name \textbf{(Ours)}     & 69.25          & \textbf{47.00}    & 56.50         & \textbf{64.75}   & 78.00           & \textbf{79.37}    & 77.71            & \textbf{54.67}    & 74.16           & \textbf{30.66}    \\ \bottomrule
\end{tabular}
}
\end{table*}

\noindent\textbf{Implementation Configuration.}
We implement our experiment on NVIDIA RTX A6000 (48GB). 
Using VLLM and the Transformers library for inference. We use the greedy decoding method for inference (set do\_sample = False).
\begin{table*}[t]
\centering
\fontsize{8pt}{6pt}\selectfont
\caption{The experimental results of sentence-level attack on 5 benchmark datasets and 6 open-source LLMs.}
\label{table:sentencelevelmain}
\resizebox{0.9\textwidth}{!}{
\begin{tabular}{cccccccccccc}
\toprule
\multirow{2}{*}{\textbf{Model}}          & \textbf{Dataset}        & \multicolumn{2}{c}{\textbf{SST-2}} & \multicolumn{2}{c}{\textbf{SMS}} & \multicolumn{2}{c}{\textbf{AGNews}} & \multicolumn{2}{c}{\textbf{DBPedia}} & \multicolumn{2}{c}{\textbf{Amazon}} \\ \cmidrule{2-12}
                                           & \textbf{Metric}       & \textbf{ACC}   & \textbf{ASR}      & \textbf{ACC}   & \textbf{ASR}    & \textbf{ACC}    & \textbf{ASR}      & \textbf{ACC}     & \textbf{ASR}      & \textbf{ACC}    & \textbf{ASR}      \\ \midrule
\multirow{6}{*}{\textbf{Llama-70B}}    & Clean              & 94.27          & 50.00             & 95.50          & 50.00           & 94.27           & 25.00             & 94.25            & 7.14              & 87.75           & 16.67             \\ \cmidrule{2-12}
                                           & ICL-1Shot             & 95.00          & 51.37             & 96.50          & 52.50           & 93.55           & 23.37             & 94.50            & 7.25              & 86.68           & 7.50              \\ \cmidrule{2-12}
                                           & ICL-2Shot             & 95.12          & 51.37             & 95.75          & 52.50           & 93.37           & 23.57             & 94.60            & 8.14              & 86.66           & 7.33              \\ \cmidrule{2-12}
                                           & ICL-3Shot             & 94.87          & 52.25             & 94.50          & 52.00           & 93.35           & 23.97             & 94.50            & 7.78              & 86.58           & 7.25              \\ \cmidrule{2-12}
                                           & \Name \textbf{(Ours)} & 95.00          & \textbf{100.00}   & 95.50          & \textbf{95.50}  & 93.85           & \textbf{100.00}   & 93.64            & \textbf{61.17}    & 86.75           & 13.25             \\ \midrule
\multirow{6}{*}{\textbf{Deepseek-7B}} & Clean              & 80.12          & 50.00             & 60.00          & 50.00           & 82.50           & 25.00             & 67.85            & 7.14              & 62.66           & 16.67             \\ \cmidrule{2-12}
                                           & ICL-1Shot             & 81.12          & 52.12             & 57.25          & 38.00           & 78.77           & 32.25             & 66.50            & 13.82             & 61.41           & 21.25             \\ \cmidrule{2-12}
                                           & ICL-2Shot             & 79.25          & 50.12             & 41.50          & 23.50           & 75.62           & 30.32             & 61.21            & 22.53             & 62.00           & 19.41             \\ \cmidrule{2-12}
                                           & ICL-3Shot             & 67.25          & 63.25             & 42.50          & 26.00           & 83.50           & 29.35             & 59.39            & 25.00             & 61.08           & 17.75             \\ \cmidrule{2-12}
                                           & \Name \textbf{(Ours)} & 82.12          & \textbf{89.75}    & 69.50          & \textbf{81.00}  & 83.10           & \textbf{96.22}    & 71.42            & \textbf{82.78}    & 62.08           & \textbf{69.58}    \\ \midrule
\multirow{6}{*}{\textbf{Llama-8B}}     & Clean              & 85.62          & 50.00             & 92.25          & 50.00           & 90.07           & 25.00             & 87.28            & 7.14              & 79.50           & 16.67             \\ \cmidrule{2-12}
                                           & ICL-1Shot             & 85.25          & 47.25             & 92.00          & 49.00           & 88.95           & 27.07             & 76.96            & 6.71              & 82.83           & 12.58             \\ \cmidrule{2-12}
                                           & ICL-2Shot             & 86.00          & 48.62             & 90.50          & 51.25           & 89.22           & 27.07             & 82.82            & 10.42             & 81.75           & 13.25             \\ \cmidrule{2-12}
                                           & ICL-3Shot             & 87.37          & 50.75             & 90.75          & 65.25           & 88.92           & 27.72             & 74.50            & 17.42             & 80.16           & 16.25             \\ \cmidrule{2-12}
                                           & \Name \textbf{(Ours)} & 88.62          & \textbf{85.50}    & 93.25          & \textbf{70.50}  & 89.62           & \textbf{95.02}    & 85.85            & \textbf{57.46}    & 81.41           & \textbf{32.08}    \\ \midrule
\multirow{6}{*}{\textbf{Yi-34B}}  & Clean              & 94.25          & 50.00             & 82.50          & 50.00           & 91.82           & 25.00             & 89.57            & 7.14              & 86.33           & 16.67             \\ \cmidrule{2-12}
                                           & ICL-1Shot             & 94.62          & 48.37             & 84.25          & 55.00           & 91.87           & 24.57             & 89.03            & 5.10              & 85.16           & 8.16              \\ \cmidrule{2-12}
                                           & ICL-2Shot             & 94.12          & 50.37             & 84.25          & 64.00           & 91.82           & 23.82             & 89.78            & 11.67             & 84.83           & 8.83              \\ \cmidrule{2-12}
                                           & ICL-3Shot             & 93.12          & 46.75             & 84.00          & 60.00           & 91.37           & 26.72             & 80.85            & 21.21             & 83.83           & 9.66              \\ \cmidrule{2-12}
                                           & \Name \textbf{(Ours)} & 94.50          & \textbf{93.87}    & 93.00          & \textbf{78.50}  & 91.67           & \textbf{94.60}    & 89.25            & \textbf{33.60}    & 85.41           & \textbf{72.91}    \\ \midrule
\multirow{6}{*}{\textbf{Mistral-Small}}    & Clean              & 91.50          & 50.00             & 77.25          & 50.00           & 87.50           & 25.00             & 71.57            & 7.14              & 81.08           & 16.67             \\ \cmidrule{2-12}
                                           & ICL-1Shot             & 83.50          & 25.75             & 74.75          & 36.50           & 84.07           & 25.50             & 52.85            & 6.46              & 73.50           & 12.33             \\ \cmidrule{2-12}
                                           & ICL-2Shot             & 84.37          & 32.25             & 72.00          & 25.75           & 83.70           & 26.00             & 51.57            & 6.50              & 75.83           & 12.58             \\ \cmidrule{2-12}
                                           & ICL-3Shot             & 83.00          & 23.25             & 71.00          & 28.00           & 84.87           & 24.95             & 23.00            & 27.85             & 58.91           & 11.58             \\ \cmidrule{2-12}
                                           & \Name \textbf{(Ours)} & 93.37          & \textbf{99.37}    & 56.75          & \textbf{57.25}  & 87.92           & \textbf{73.65}    & 75.53            & \textbf{73.89}    & 81.16           & 5.66              \\ \midrule
\multirow{6}{*}{\textbf{Phi-3.5}}          & Clean              & 68.50          & 50.00             & 57.00          & 50.00           & 81.72           & 25.00             & 77.50            & 7.14              & 75.16           & 16.67             \\ \cmidrule{2-12}
                                           & ICL-1Shot             & 6.12           & 0.87              & 56.75          & 49.50           & 59.62           & 31.37             & 45.03            & 7.07              & 55.41           & 4.58              \\ \cmidrule{2-12}
                                           & ICL-2Shot             & 8.25           & 3.87              & 50.25          & 57.75           & 61.02           & 34.15             & 49.82            & 11.85             & 53.50           & 3.33              \\ \cmidrule{2-12}
                                           & ICL-3Shot             & 16.62          & 11.37             & 51.25          & 56.75           & 57.07           & 39.47             & 50.53            & 22.25             & 57.83           & 5.58              \\ \cmidrule{2-12}
                                           & \Name \textbf{(Ours)} & 65.62          & \textbf{77.37}    & 58.50          & 55.00           & 77.97           & \textbf{80.10}    & 76.39            & \textbf{28.75}    & 72.91           & \textbf{20.66}    \\ \bottomrule
\end{tabular}
}
\end{table*}
\section{Experimental Results}\label{subsection:experimentalresult}


\subsection{Results on Open-source LLMs}\label{subsubsection:Wordlevelresult}
\noindent\textbf{Word-level Attack.}
As shown in \autoref{table:wordlevelmain}, the following key observations can be summarized as follows. 
First, a high ASR is achieved across most models and datasets, which is substantially higher than that of the ICL backdoor.
Specifically, on the SMS dataset, Llama-70B, Deepseek-7B, Llama-8B, Yi-34B, Mistral-Small, and Phi-3.5 attain ASRs of 100.00\%, 96.25\%, 98.75\%, 74.00\%, 100\%, 47.00\%, respectively. 
These values are substantially higher than those of the ICL-3shot backdoor, which attain ASRs of 52.25\%, 66.87\%, 48.75\%, 47.50\%, 35.25\%, and 8.62\%, respectively.
Moreover, as the class number of the dataset increases, the ASR tends to decrease.
It can be observed that the ASR of the two binary classification datasets, SST-2 and SMS, is slightly higher than that of the other three multi-class datasets.
In conclusion, word-level attack achieves high effectiveness compared with baseline methods while significantly maintaining clean performance, and as the number of classes increases, the attack performance decreases overall.\looseness=-1

\noindent\textbf{Sentence-level Attack.}\label{subsubsection:Sentencelevelresult}
The experimental results can be found in the \autoref{table:sentencelevelmain}.
The following crucial points can be summarized as follows.
Sentence-level attack achieves extremely high ASR compared with baselines. 
Specifically, the ASR values on the Llama-70B, Llama-8B, Deepseek-7B, Yi-34B, Mistral-Small, and Phi-3.5 models with the SST-2 dataset are 100\%, 89.75\%, 85.50\%, and 93.87\%, respectively. 
However, in rare cases, the ASR is relatively low.
We speculate that this is due to the model's limited ability.
For some datasets that have longer sentences, insufficient semantic understanding and instruction-following abilities of the model prevent the backdoor instruction from taking effect, which will be proved in \autoref{subsubsection:closellm}.
Meanwhile, the ACC shows a minimal decrease.
For instance, on the Llama-70B model and the SST-2 dataset, the ACC of sentence-level attack increases by 0.73\%. 
Moreover, compared to the word-level attack, the sentence-level attack achieves a slightly lower ASR, but it will achieve a more stealthy attack following the previous conclusion~\cite{backdoorsurvey}.
In conclusion, the sentence-level attack achieves an effective attack compared to the baselines while maintaining clean performance.

\subsection{Results on Closed-source LLMs}\label{subsubsection:closellm}
To demonstrate that, as LLMs’ instruction-following ability improves, the attack effectiveness of \Name also increases, we conduct extensive experiments on 5 datasets using 3 state-of-the-art closed-source LLMs. 

\noindent\textbf{Word-level Attack.} 
As shown in \autoref{table:closeword}, we draw the following conclusions.
First, \Name achieves a strong attack performance at the word-level attack in closed-source LLMs, extremely higher than baselines. 
Specifically, on the SST-2 dataset with Gemini-2.5, it achieves a 100\% ASR while ACC exhibits almost no degradation relative to the baseline, which is higher than 59.25\% of ICL-3Shot. 
Second, the ASR on closed-source commercial LLMs is substantially higher than that on open-source LLMs. 
Concretely, on Gemini-2.5, we obtain ASR of 99.83\% for word-level attacks, respectively, whereas on Yi-34B, the corresponding rates are 47.25\%. 
We speculate that this outcome arises because closed-source LLMs possess stronger instruction-following capabilities and therefore follow the backdoor instruction more reliably, yielding higher ASR. 
This observation indicates that \Name exhibits strong adaptability as LLM's instruction-following capabilities improve. 
In conclusion, we achieve superior attack performance on closed-source LLMs than on open-source LLMs, which demonstrates the developmental adaptability of \Name. \looseness=-1

\noindent\textbf{Sentence-level Attack.} 
As shown in \autoref{table:closesentence}, we draw similar conclusions in the word-level attack.
First, \Name still achieves a stronger attack performance at the sentence-level attack in closed-source LLMs than baselines. 
Second, the ASR on closed-source commercial LLMs is also substantially higher than that on open-source LLMs, which is also because of the stronger instruction-following capabilities in closed-source commercial LLMs.  \looseness=-1

\begin{table*}[t]
\centering
\fontsize{8pt}{6pt}\selectfont

\caption{The experimental results of the word-level attack on 5 benchmark datasets and 3 closed-source LLMs.}\label{table:closeword}
\resizebox{0.9\textwidth}{!}{
\begin{tabular}{cccccccccccc}
\toprule 
\multirow{2}{*}{\textbf{Model}}                             & \textbf{Dataset}  & \multicolumn{2}{c}{\textbf{SST-2}} & \multicolumn{2}{c}{\textbf{SMS}} & \multicolumn{2}{c}{\textbf{AGNews}} & \multicolumn{2}{c}{\textbf{DBPedia}} & \multicolumn{2}{c}{\textbf{Amazon}} \\ \cmidrule{2-12}
                    & \textbf{Metric} & \textbf{ACC}   & \textbf{ASR}      & \textbf{ACC}  & \textbf{ASR}     & \textbf{ACC}    & \textbf{ASR}      & \textbf{ACC}    & \textbf{ASR}       & \textbf{ACC}    & \textbf{ASR}      \\ \midrule
\multirow{6}{*}{\textbf{Gemini-2.5}}  & Clean        & 94.75          & 50.00             & 95.25         & 50.00            & 93.32           & 25.00             & 94.25           & 7.14               & 88.25           & 16.67             \\ \cmidrule{2-12}
                                      & ICL-1Shot       & 95.25          & 52.50             & 94.75         & 54.50            & 92.87           & 23.82             & 94.53           & 9.28               & 87.66           & 13.33             \\ \cmidrule{2-12}
                                      & ICL-2Shot       & 94.25          & 52.50             & 92.00         & 61.50            & 92.65           & 24.02             & 92.82           & 12.85              & 87.58           & 12.41             \\ \cmidrule{2-12}
                                      & ICL-3Shot       & 92.25          & 59.25             & 92.75         & 60.00            & 92.17           & 27.70             & 82.28           & 32.71              & 85.41           & 19.00             \\ \cmidrule{2-12}
                                      & \Name \textbf{(Ours)}      & 94.62          & \textbf{100.00}   & 96.25         & \textbf{100.00}  & 93.30           & \textbf{100.00}   & 93.89           & \textbf{99.85}     & 88.41           & \textbf{99.83}    \\ \midrule
\multirow{6}{*}{\textbf{GPT-4o-mini}} & Clean        & 92.37          & 50.00             & 96.25         & 50.00            & 91.47           & 25.00             & 91.46           & 7.14               & 88.50           & 16.67             \\ \cmidrule{2-12}
                                      & ICL-1Shot       & 94.75          & 53.62             & 94.50         & 55.75            & 91.35           & 23.77             & 93.00           & 9.14               & 88.83           & 10.08             \\ \cmidrule{2-12}
                                      & ICL-2Shot       & 94.37          & 53.37             & 93.00         & 56.50            & 91.22           & 23.97             & 92.92           & 9.25               & 88.50           & 10.50             \\ \cmidrule{2-12}
                                      & ICL-3Shot       & 94.25          & 53.12             & 93.25         & 56.25            & 91.75           & 23.65             & 92.67           & 9.53               & 89.33           & 11.66             \\ \cmidrule{2-12}
                                      & \Name \textbf{(Ours)}      & 92.25          & \textbf{100.00}   & 94.00         & \textbf{100.00}  & 91.42           & \textbf{100.00}   & 90.25           & \textbf{100.00}    & 87.75           & \textbf{92.50}    \\ \midrule
\multirow{6}{*}{\textbf{Deepseek-v3.2}}    & Clean        & 95.25          & 50.00             & 95.75         & 50.00            & 93.42           & 25.00             & 94.89           & 7.14               & 89.16           & 14.25             \\ \cmidrule{2-12}
                                      & ICL-1Shot       & 95.50          & 52.50             & 96.50         & 50.50            & 92.22           & 24.37             & 95.07           & 7.89               & 88.50           & 10.58             \\ \cmidrule{2-12}
                                      & ICL-2Shot       & 95.50          & 52.50             & 95.50         & 49.25            & 92.32           & 24.22             & 94.64           & 8.82               & 88.91           & 9.91              \\ \cmidrule{2-12}
                                      & ICL-3Shot       & 94.50          & 56.62             & 95.75         & 50.50            & 91.85           & 25.17             & 93.50           & 10.71              & 89.58           & 13.58             \\ \cmidrule{2-12}
                                      & \Name \textbf{(Ours)}      & 93.37          & \textbf{100.00}   & 91.75         & \textbf{100.00}  & 89.70           & \textbf{100.00}   & 94.03           & \textbf{99.96}     & 89.71           & \textbf{97.66}    \\ \bottomrule

\end{tabular}
}
\end{table*}

\begin{table*}[t]
\fontsize{8pt}{6pt}\selectfont

\centering
\caption{The experimental results of the sentence-level attack on 5 benchmark datasets and 3 closed-source LLMs.}\label{table:closesentence}
\resizebox{0.9\textwidth}{!}{
\begin{tabular}{cccccccccccc}
\toprule
\multirow{2}{*}{\textbf{Model}}                             & \textbf{Dataset}  & \multicolumn{2}{c}{\textbf{SST-2}} & \multicolumn{2}{c}{\textbf{SMS}} & \multicolumn{2}{c}{\textbf{AGNews}} & \multicolumn{2}{c}{\textbf{DBPedia}} & \multicolumn{2}{c}{\textbf{Amazon}} \\ \cmidrule{2-12}
                     & \textbf{Metric} & \textbf{ACC}   & \textbf{ASR}      & \textbf{ACC}  & \textbf{ASR}     & \textbf{ACC}    & \textbf{ASR}      & \textbf{ACC}    & \textbf{ASR}       & \textbf{ACC}    & \textbf{ASR}      \\ \midrule
\multirow{6}{*}{\textbf{Gemini-2.5}}  & Clean        & 94.75          & 50.00             & 95.25         & 50.00            & 93.32           & 25.00             & 94.25           & 7.14               & 88.25           & 16.67             \\ \cmidrule{2-12}
                                      & ICL-1Shot       & 95.50          & 53.12             & 95.75         & 55.75            & 92.77           & 24.05             & 94.32           & 7.82               & 88.33           & 14.25             \\ \cmidrule{2-12}
                                      & ICL-2Shot       & 95.37          & 53.62             & 94.00         & 60.00            & 93.20           & 23.87             & 94.14           & 10.78              & 87.83           & 13.91             \\ \cmidrule{2-12}
                                      & ICL-3Shot       & 91.75          & 58.25             & 95.25         & 58.00            & 92.80           & 25.02             & 86.50           & 37.42              & 81.75           & 31.91             \\ \cmidrule{2-12}
                                      & \Name \textbf{(Ours)}  & 95.75          & \textbf{100.00}   & 95.50         & \textbf{100.00}  & 93.32           & \textbf{100.00}   & 93.75           & \textbf{100.00}    & 88.75           & \textbf{100.00}   \\ \midrule
\multirow{6}{*}{\textbf{GPT-4o-mini}} & Clean        & 92.37          & 50.00             & 96.25         & 50.00            & 91.47           & 25.00             & 91.46           & 7.14               & 88.50           & 16.67             \\ \cmidrule{2-12}
                                      & ICL-1Shot       & 94.62          & 53.75             & 93.75         & 54.00            & 90.87           & 24.02             & 93.10           & 9.21               & 89.16           & 10.83             \\ \cmidrule{2-12}
                                      & ICL-2Shot       & 94.62          & 53.62             & 95.50         & 53.50            & 90.77           & 24.15             & 93.03           & 9.64               & 88.75           & 10.25             \\ \cmidrule{2-12}
                                      & ICL-3Shot       & 94.00          & 54.37             & 94.25         & 53.75            & 91.65           & 23.60             & 92.57           & 9.71               & 89.00           & 11.00             \\ \cmidrule{2-12}
                                      & \Name \textbf{(Ours)}  & 95.75          & \textbf{100.00}   & 96.50         & \textbf{100.00}  & 91.80           & \textbf{100.00}   & 91.85           & \textbf{100.00}    & 88.50           & \textbf{91.16}    \\ \midrule
\multirow{6}{*}{\textbf{Deepseek-v3.2}}    & Clean        & 95.25          & 50.00             & 95.75         & 50.00            & 93.42           & 25.00             & 94.89           & 7.14               & 89.16           & 14.25             \\ \cmidrule{2-12}
                                      & ICL-1Shot       & 94.87          & 55.25             & 96.75         & 50.75            & 92.35           & 24.82             & 94.53           & 8.17               & 88.91           & 10.41             \\ \cmidrule{2-12}
                                      & ICL-2Shot       & 94.50          & 55.62             & 96.75         & 50.25            & 92.47           & 24.32             & 94.25           & 9.57               & 88.66           & 10.08             \\ \cmidrule{2-12}
                                      & ICL-3Shot       & 94.87          & 55.87             & 96.25         & 50.75            & 93.32           & 25.30             & 93.07           & 11.32              & 89.08           & 11.00             \\ \cmidrule{2-12}
                                      & \Name \textbf{(Ours)}  & 89.87          & \textbf{100.00}   & 95.75         & \textbf{100.00}  & 92.25           & \textbf{100.00}   & 94.32           & \textbf{99.92}     & 89.16           & \textbf{97.50}    \\ \bottomrule
\end{tabular}
}
\end{table*}

\subsection{Ablation Study}\label{subsection:ablationstudy}
In the ablation study, the Llama-70B~\cite{llama3paper} model and the five datasets adopted in the main experiment are employed for the experiments.
The remaining experimental settings are the same as those in the main experiment.

\begin{figure}[t]
    \centering
    \includegraphics[width = 0.45\textwidth]{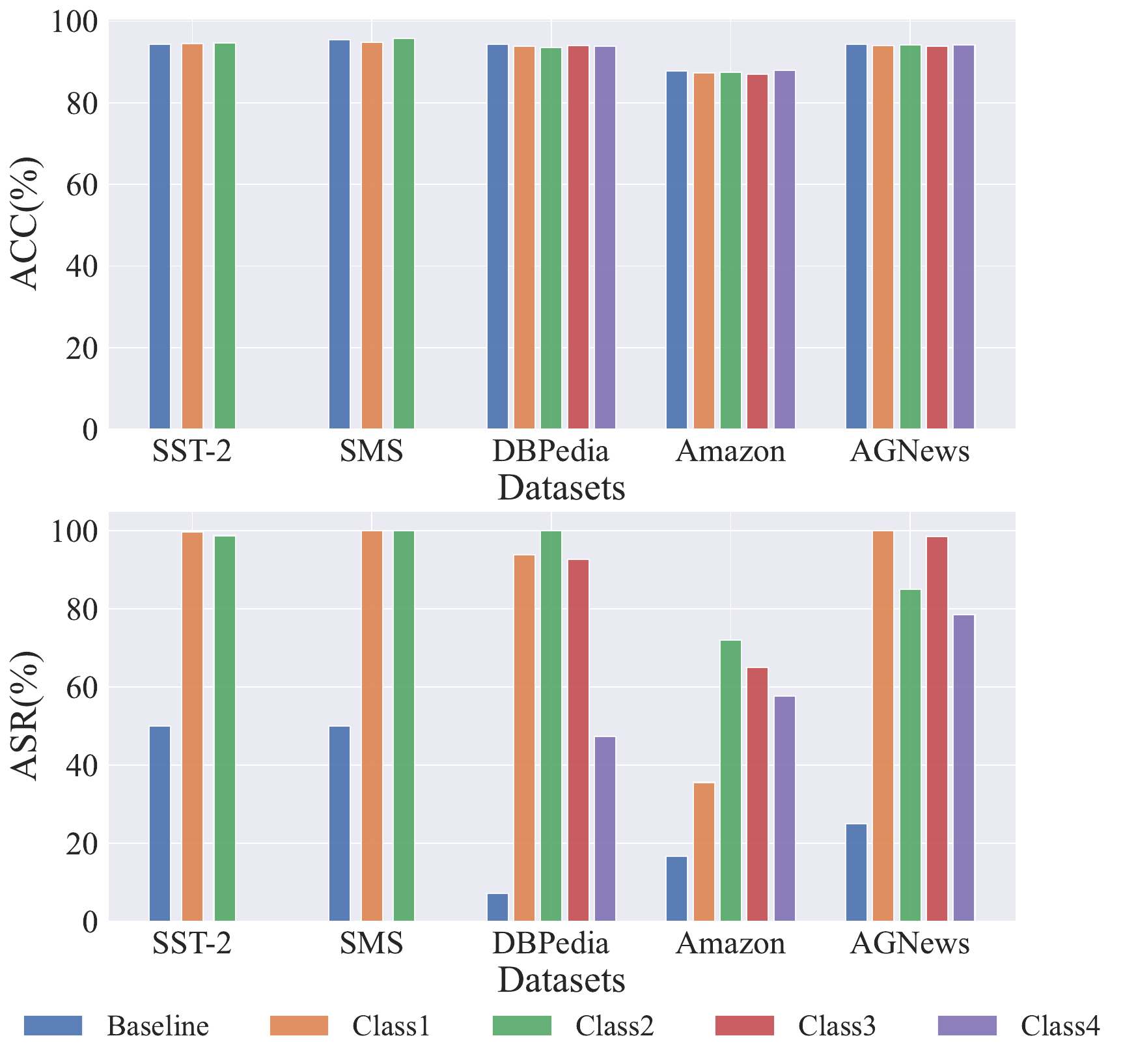}
    \caption{Evaluating the effect of different target label on Llama-70B~\cite{llama3paper} utilizing 5 datasets in the main experiment.}\label{fig:label}
\end{figure}

\begin{figure}[t]
    \centering
    \includegraphics[width = 0.5\textwidth]{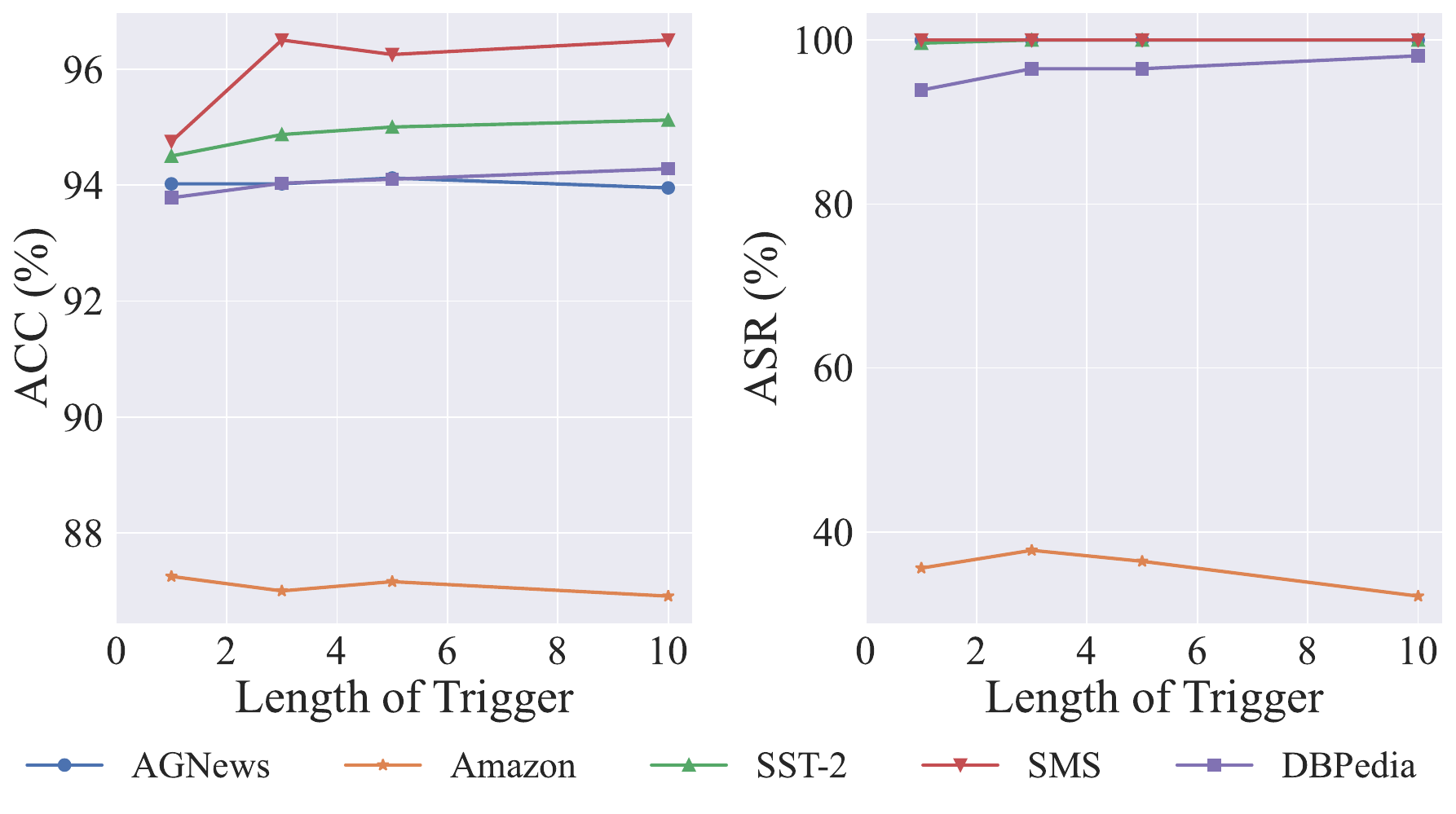}
    \caption{Evaluating the effect of the length of trigger on Llama-70B~\cite{llama3paper} utilizing 5 datasets in the main experiment.}\label{fig:lengthoftrigger}
\end{figure}

\noindent\textbf{Different Labels.}
We aim to comprehensively evaluate the effectiveness of \Name across different target labels. 
To this end, we select the first four classes of AGNews, Amazon, and DBPedia as attack targets. Because SST-2 and SMS constitute binary classification tasks, we select only their first two classes for evaluation.
The experimental results appear in \autoref{fig:label}, where Class i denotes the i-th class. 
We summarize our findings as follows:
First, under various target-label settings, the variation of ACC is small and in some cases even exceeds the baseline.
Second, under a few target-label settings, ASR exhibits certain fluctuations, but overall it remains substantially higher than the baseline.
Concretely, on the SMS dataset, ASR almost always remains at 100\%; on the DBPedia dataset, ASR approaches 100\% for Classes 1, 2, and 3, whereas ASR for Class 4 is only about 50\%. 
We speculate that this discrepancy stems from the LLM’s limited knowledge and representation of samples in that class, which attenuates its responsiveness to the backdoor instruction inserted by \Name in the system prompt.
In summary, across different target label configurations, the clean performance remains essentially stable, while attack effectiveness decreases only in a small number of cases. \looseness=-1

\noindent\textbf{Length of The Trigger.}
We investigate the trigger length in the word-level attack, where we expand the trigger length by repeating the \textit{cf} string.
The \textit{cf} string is repeated 3, 5, and 10 times. The experimental results are shown in \autoref{fig:lengthoftrigger}. 
The following findings can be summarized.
The ACC of most datasets increases with the trigger length.
Specifically, in the ACC subplot for SMS, SST-2, and DBPedia, this trend is more pronounced.
We speculate that as the trigger length increases, it appears with a lower frequency in the input, and is less likely to be triggered by words resembling triggers within some normal words, which leads to a consequently higher ACC.
However, the trend for ASR with increasing trigger length is not typical.
For example, in the ASR subplot, the Amazon dataset shows an increasing-then-decreasing trend, while the DBPedia dataset exhibits a consistent increase. We speculate that this variation is due to the significant differences in data distributions across different datasets, leading to greater randomness.
In summary, as the trigger length increases, the clean performance increases, while the trend of attack effectiveness is not typical.

\begin{figure}[t]
    \centering
    \includegraphics[width = 0.85\columnwidth]{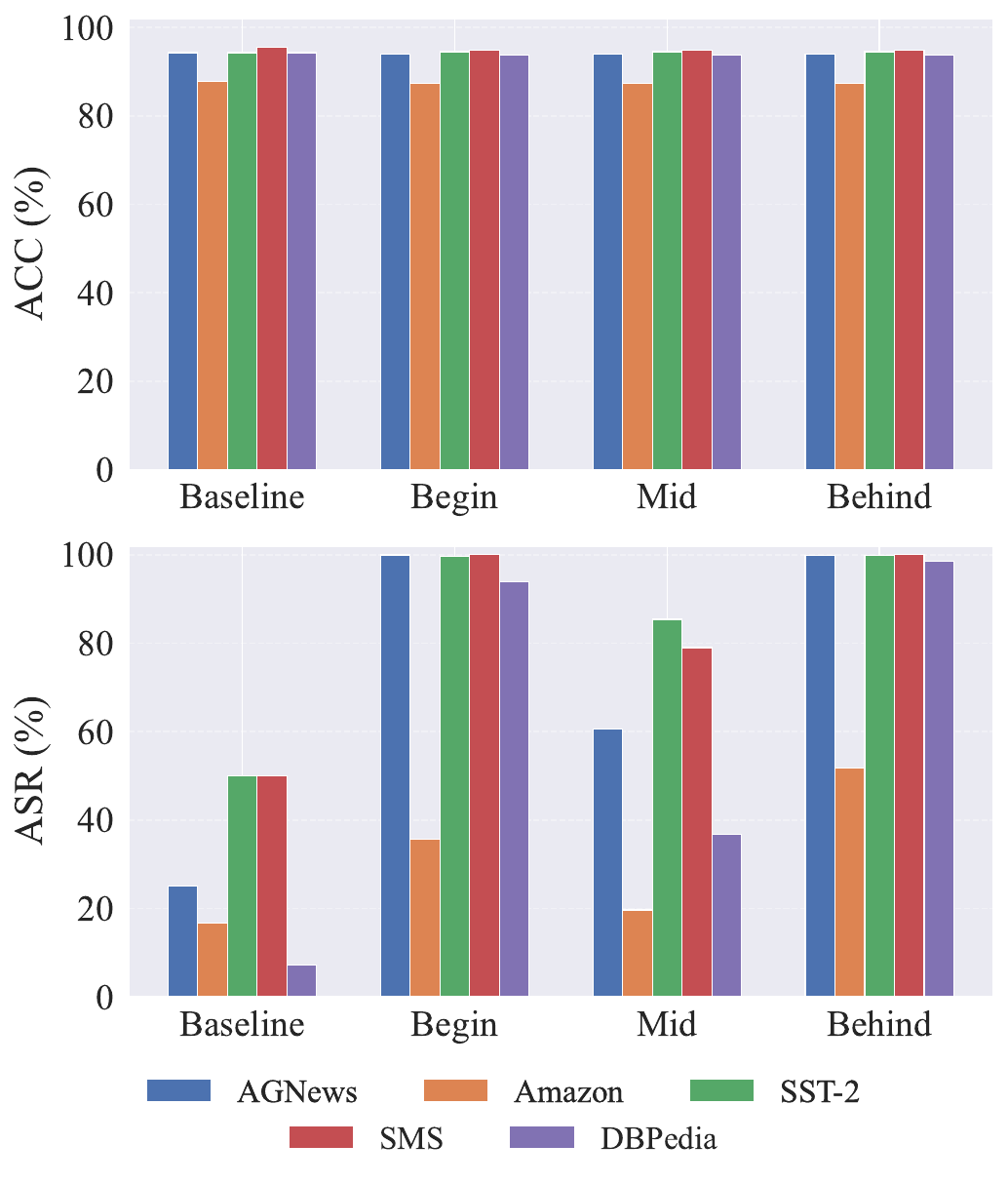}
    \caption{Evaluating the effect of position of the trigger on Llama-70B~\cite{llama3paper} utilizing 5 datasets in the main experiment. Beginning means backdoor instruction is at the beginning of the sentence and so on.}\label{fig:positionoftrigger}
\end{figure}
\noindent\textbf{Position of The Trigger.}
We investigate the position of the trigger in the word-level attack, with the trigger placed at the beginning, middle, and end of the sentence for the experiment. 
The results are shown in \autoref{fig:positionoftrigger}.
Overall, the change in ACC with respect to the trigger position is minimal.
However, the ASR is the highest when the trigger is at the end of the sentence, while it is the lowest when the trigger is in the middle.
The findings are summarized as follows: The change in ACC is small across different positions. 
As shown in the ACC subplot, there is little difference between the beginning, middle, and end positions across the five test datasets. 
The ACC in all three positions is almost identical to the baseline. 
The performance of the beginning, middle, and end is nearly the same as the baseline across the five test datasets.
The ASR is the highest when the trigger is at the end of the sentence, while it is the lowest when the trigger is in the middle, as shown in the ASR subplot. The end position consistently achieves the best performance across the five datasets, with the beginning being very close to the end.
In contrast, the middle shows poor performance, achieving the worst results across all datasets. 
This may be because the trigger features at the beginning and end of the sentence are more prominent, such that easier for the model to recognize, thereby exhibiting an effective backdoor, while features in the middle are more latent and harder for the model to identify.
In summary, the position of the trigger does not affect the performance on clean samples, while inserting the trigger at the end of the sentence yields the highest attack effectiveness, and inserting it in the middle results in the lowest effectiveness.

\begin{figure}[t]
    \centering
    \includegraphics[width = 1\columnwidth]{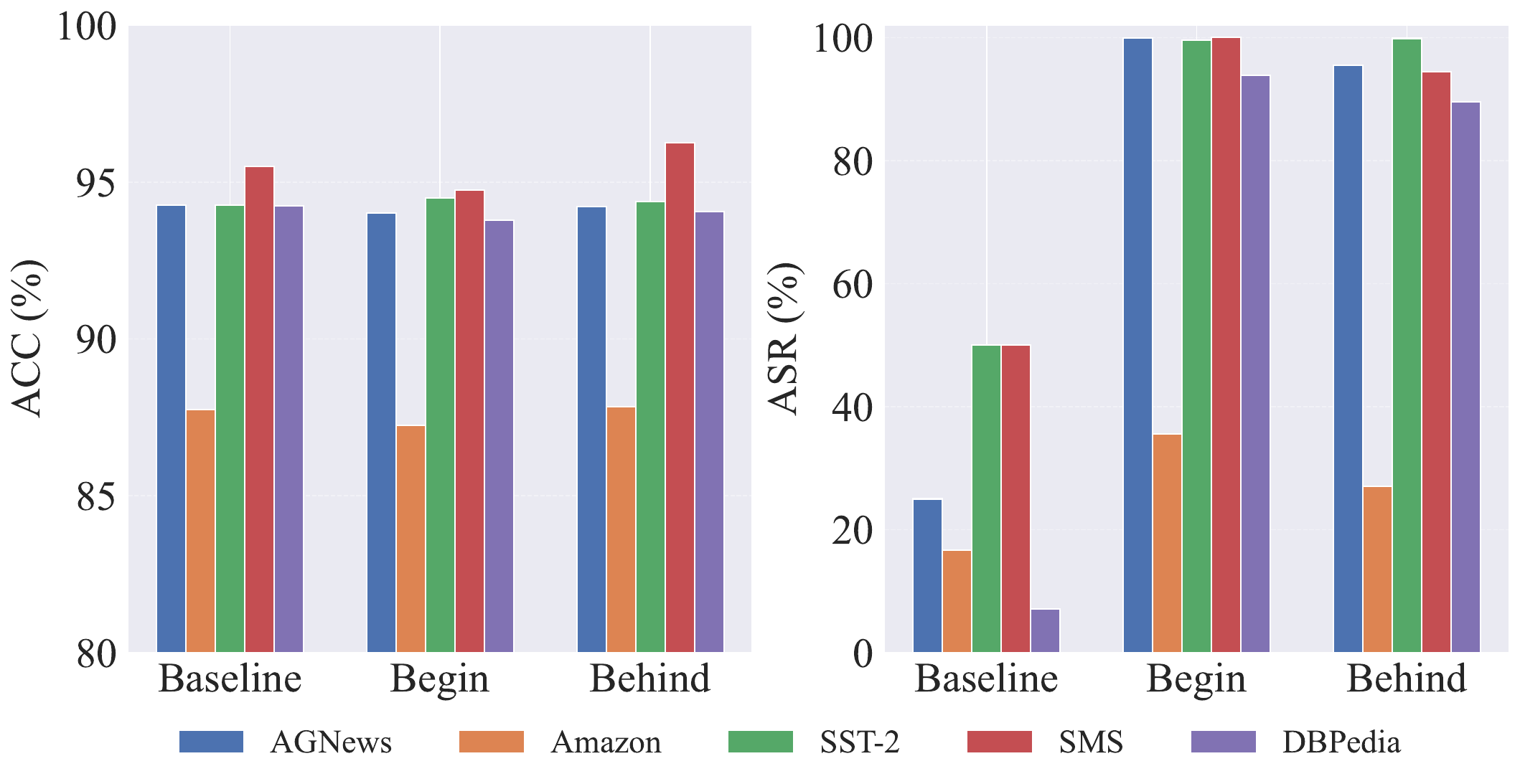}
    \caption{Evaluating the effect of position of the backdoor instruction on Llama-70B~\cite{llama3paper} utilizing 5 datasets in the main experiment. Beginning means backdoor instruction is at the beginning of the system prompt and so on.}\label{fig:positionofinst}
\end{figure}
\begin{figure}[t]
    \centering
    \includegraphics[width = 1\columnwidth]{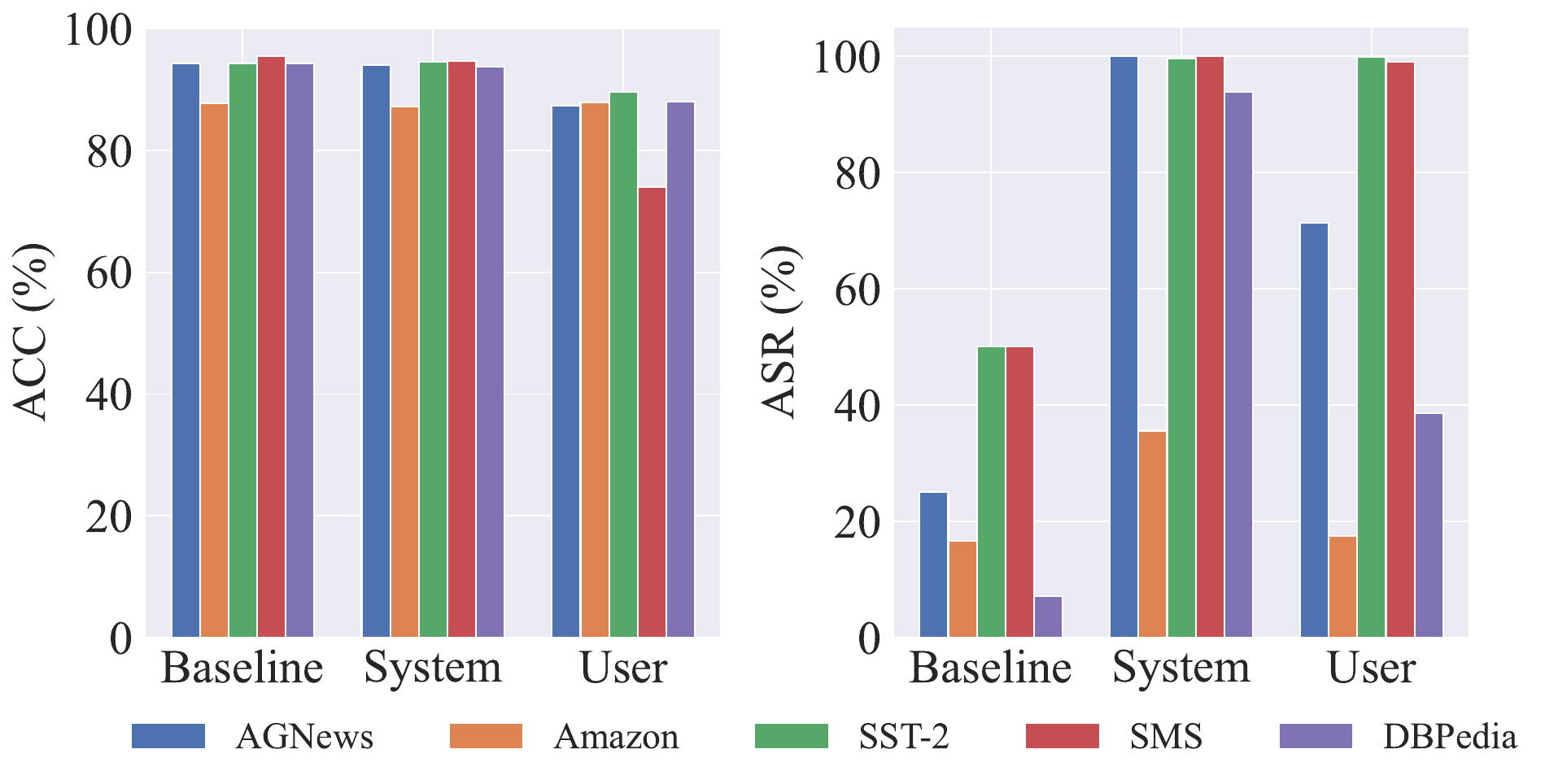}
    \caption{Evaluating the effect of different roles the backdoor instruction in on Llama-70B\cite{llama3paper} using 5 datasets in the main experiment.}\label{fig:roleoftrigger}
\end{figure}
\noindent\textbf{Position of The Backdoor Instruction.}
The relative positional relationship between backdoor instruction and task instruction in the system prompt of the word-level attack is investigated.
The beginning means the start of the sentence, before the task instruction, while the end refers to the position following the task instruction.
The experimental results are presented in \autoref{fig:positionofinst}. 
The following conclusions can be drawn.
The position of the end slightly outperforms the beginning position in maintaining the ACC metric.
Specifically, in 4 out of 5 datasets, the ACC of the end position is slightly higher than that of the beginning position.
However, in terms of the ASR metric, the beginning position performs better than the end position. 
Specifically, on all five datasets, the ASR of the beginning position is higher than that of the end position.
In summary, the attack is more effective when the backdoor instruction is at the beginning of the system prompt, while the clean performance is slightly higher when the backdoor instruction is at the end of the system prompt.

\noindent\textbf{Roles The Backdoor Instruction in.}
The experimental results are shown in the \autoref{fig:roleoftrigger}.
Overall, the backdoor instruction exhibits higher ACC and ASR in the \textit{system} role compared to the \textit{user} role.
The following conclusions can be summarized.
The ASR of the \textit{system} is generally higher than that of the \textit{user}.
In the ASR subplot, it is observed that the \textit{system} outperforms the \textit{user} across all five datasets. 
Furthermore, the ACC of the \textit{system} is higher than that of the \textit{user} in most cases.
In the ACC subplot, it is observed that the \textit{system} outperforms the \textit{user} in four datasets, except the Amazon dataset, where the performance of the \textit{system} is slightly lower than that of the \textit{user}.
We speculate it is because the system prompt has a higher priority in the control of LLMs.
This difference is attributed to the higher priority of the \textit{system} role's instructions compared to those of the \textit{user} role, resulting in a higher attack effectiveness in the \textit{system} role.
In summary, the attack effectiveness and clean performance when backdoor instruction is used in the \textit{system} role are better than in the \textit{user} role.
\begin{figure}[t]
    \centering
    \includegraphics[width = 1\columnwidth]{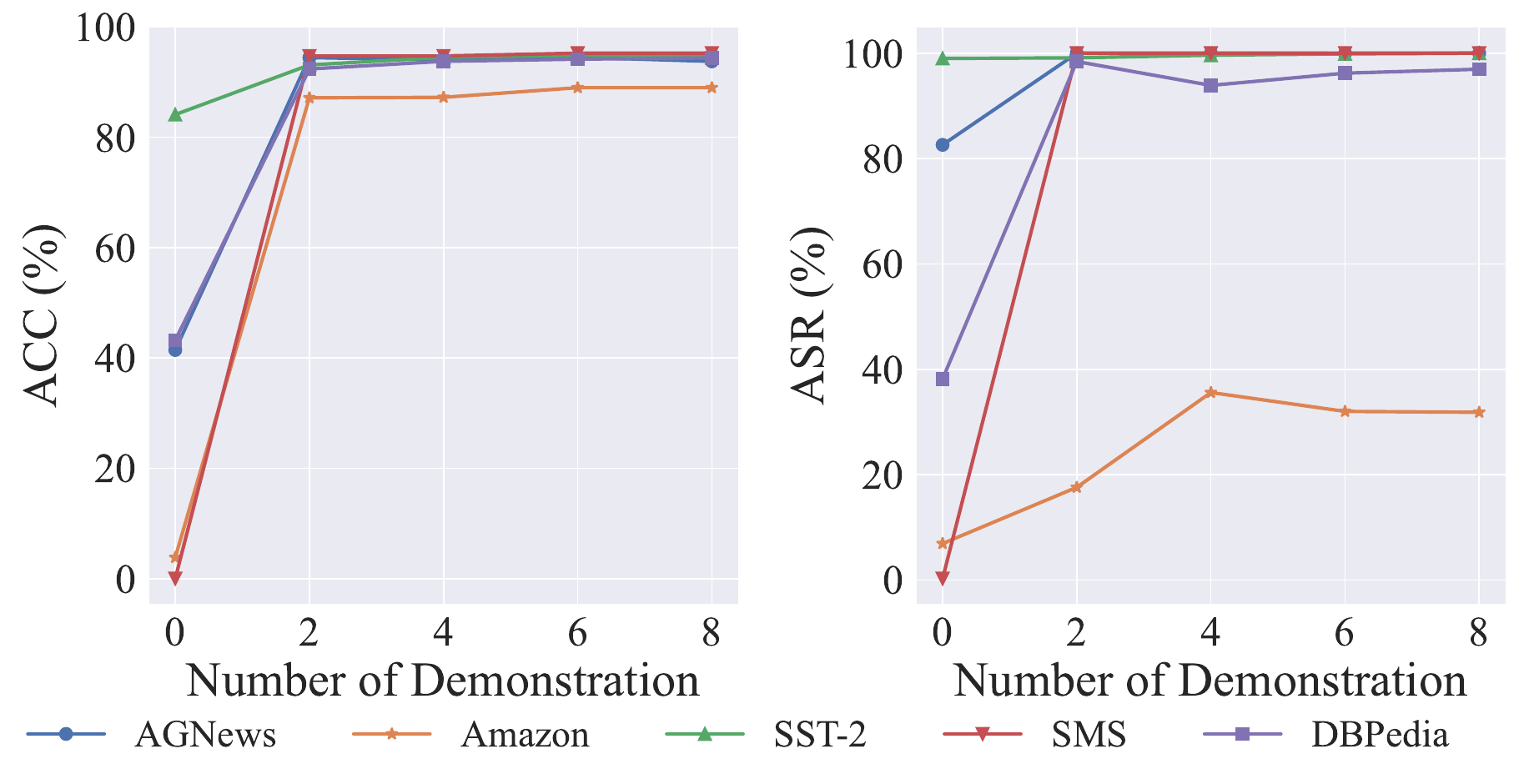}
    \caption{Evaluating the effect of number of demonstrations on Llama-70B~\cite{llama3paper} utilizing 5 datasets in the main experiment.}\label{fig:triggerdemon}
\end{figure}

\noindent\textbf{Number of Demonstrations.}
We conduct an ablation experiment on the number of demonstrations used during inference introduced in \autoref{subsection:overalldesign}. 
The results are shown in \autoref{fig:triggerdemon}. 
We can draw the following conclusions as follows.
When the number of demonstrations increases from 0 to 2, both ACC and ASR show a strong upward trend as shown in the plots.
We speculate that it is because when utilizing 0 demonstrations, the model has no prior knowledge of the task and fails to generalize well to the domain. 
Without constraints of demonstration, the model may generate the answer not in the label space, resulting in lower ACC and ASR.
However, the upward trend from 2 to 8 demonstrates a smaller improvement. We observe a slight increase in ASR and ACC between 2 and 4, but no noticeable improvement from 4 to 8. 
We speculate that two demonstrations are already sufficient for effective generalization to the target domain. Therefore, additional demonstrations do not significantly enhance ACC or ASR.
In summary, 2 demonstrations are enough for the \Name attack. 
Among these, the attack effectiveness is best when the number of demonstrations is 4 and 6.

\section{Discussion}\label{section:discussion}



\begin{table}[t]
\fontsize{8pt}{6pt}\selectfont
\caption{The experimental results of the LLM-as-a-judge defense on 5 benchmark datasets and 4 LLMs utilizing GPT-4o. 
Word and Sentence in the table mean word-level and sentence-level attack, respectively.}
\label{table:defense}
\resizebox{0.5\textwidth}{!}{
\begin{tabular}{ccccccc}
\toprule
                                    & \textbf{Dataset} & \textbf{SST-2} & \textbf{AGNews} & \textbf{SMS} & \textbf{DBPedia} & \textbf{Amazon} \\ \midrule
\textbf{Model}                      & \textbf{Metric}  & \textbf{TPR}   & \textbf{TPR}    & \textbf{TPR} & \textbf{TPR}     & \textbf{TPR}    \\ \midrule
\multirow{3}{*}{\textbf{Llama-8B}}  & Clean         & 0.00           & 0.00            & 0.00         & 0.00             & 0.00            \\ \cmidrule{2-7}
                                    & Word             & 0.00           & 80.00           & 0.00         & 0.00             & 0.00            \\ \cmidrule{2-7}
                                    & Sentence         & 0.00           & 0.00            & 0.00         & 0.00             & 80.00           \\ \midrule
\multirow{3}{*}{\textbf{Deepseek-7B}} & Clean        & 0.00           & 0.00            & 0.00         & 0.00             & 0.00            \\ \cmidrule{2-7}
                                    & Word             & 0.00           & 80.00           & 0.00         & 0.00             & 0.00            \\ \cmidrule{2-7}
                                    & Sentence         & 0.00           & 0.00            & 0.00         & 0.00             & 20.00           \\ \midrule
\multirow{3}{*}{\textbf{Llama-70B}} & Clean         & 0.00           & 0.00            & 0.00         & 0.00             & 0.00            \\ \cmidrule{2-7}
                                    & Word             & 0.00           & 0.00            & 0.00         & 0.00             & 0.00            \\ \cmidrule{2-7}
                                    & Sentence         & 0.00           & 0.00            & 0.00         & 0.00             & 0.00            \\ \midrule
\multirow{3}{*}{\textbf{Yi-34B}}  & Clean         & 0.00           & 0.00            & 0.00         & 0.00             & 0.00            \\ \cmidrule{2-7}
                                    & Word             & 0.00           & 0.00            & 0.00         & 0.00             & 0.00            \\ \cmidrule{2-7}
                                    & Sentence         & 0.00           & 0.00            & 0.00         & 0.00             & 40.00           \\ \bottomrule
\end{tabular}
}
\end{table}

\subsection{Potiential Defense}

Due to the invisibility of \Name without checking the chat template, the best defense is to manually or automatically verify whether there are any malicious instructions in the chat template within the tokenizer in the third-party platform.
However, our experiments demonstrate that the detection mechanisms of Hugging Face~\cite{huggingface}, the most widely used third-party platform, entirely fail to detect \Name, which leaves a potential vulnerability that \Name can exploit.
Furthermore, we attempt to mitigate \Name. Following previous studies, we employ the LLM-as-a-judge approach to detect whether the chat templates of the target LLMs contain malicious instructions~\cite{llmasajudge}.
Specifically, we use the GPT-4o\cite{openai2024gpt4ocard}, the state-of-the-art LLM, following a detection instruction that evaluates whether a given chat template contains harmful instructions.
The details of the detection instruction are shown in the Appendix.
The detection performance is measured by the True Positive Rate (\textbf{TPR}), which represents the ratio of chat templates that are identified as malicious by GPT-4o.
We conduct experiments on word-level and sentence-level attacks for four models: Llama-8B, DeepSeek-7B, Llama-70B, and Yi-34B across five benchmark datasets in the main experiment. 
Each chat template is queried five times to mitigate the randomness of LLM outputs. \looseness=-1

As shown in the experimental results \autoref{table:defense}, several key observations can be drawn.
First, the vast majority of chat templates generated by both word-level and sentence-level attacks are not detected as malicious by GPT-4o, especially for the SST-2, SMS, and DBPedia datasets, where all of the TPRs are 0.
This indicates that \Name remains stealthy under the LLM-as-a-judge~\cite{llmasajudge} defense.
Second, in a few cases on the AGNews and Amazon datasets, the TPR increases.
For example, the TPR reaches 80\% for the sentence-level attack on Llama-8B with the Amazon dataset.
We hypothesize that this occurs because the backdoor target classes in these datasets (e.g., World, Health Care) are less likely to appear as common words in typical chat templates, making them easier for the model to identify.
In conclusion, the results demonstrate that \Name achieves a high degree of stealthiness against LLM-as-a-judge defenses.
Therefore, we strongly call for novel and effective defenses for respective third-party platforms and users to detect \Name attacks. \looseness=-1

\subsection{Boarder Impact}
To eliminate potential ambiguity, this paper defines \Name as a training-free backdoor attack in which an adversary injects malicious backdoor instructions into chat templates. This manipulation alters the output behavior of an LLM without modifying its parameters or retraining the model.
Essentially, \Name belongs to the broader class of prompt injection attacks that hijack model behavior by embedding malicious instructions into inputs~\cite{promptinjectionsurvey1,promptinjectionsurvey2,owaspPI}. Such attacks can lead to a wide range of harms, including but not limited to preference manipulation~\cite{mpma}, sensitive information leakage~\cite{promptinjectionprivacy}, and execution of malicious commands~\cite{promptinjectionmaliciouscode}.
Unlike conventional prompt injection, where attackers can typically embed instructions only within user inputs, \Name uniquely targets and manipulates the system prompt. 
Since the system prompt has a higher decoding priority and generally remains persistent across sessions~\cite{systemprompt}, \Name poses a significantly greater threat than attacks confined to a single user input.
\looseness=-1



\section{Conclusion}
This paper takes the first step to reveal the vulnerability of the chat template for prompt-based backdoor attacks. 
We propose a training-free prompt-based backdoor attack, named \Name. 
\Name works by modifying the chat template in the tokenizer while maintaining the parameters of LLMs unchanged. 
Therefore, \Name allows backdoor instructions to be inserted into the input without the user's knowledge, thereby executing the prompt-based backdoor.
Based on this, the study introduces two attack variants, including the word-level attack and sentence-level attack. 
The former has better effectiveness, while the latter is distinguished by better stealthiness.
Extensive experiments are conducted on 6 popular open-source and 3 state-of-the-art closed-source LLMs across 5 benchmark datasets, with results demonstrating the effectiveness of the proposed attack. 
The word-level attack and sentence-level attack achieve up to 100\% ASR. 
The introduction of \Name highlights the security in the LLM supply chain and aims to promote the development of relevant defenses. \looseness=-1





\bibliographystyle{plain}
\bibliography{references}

\begin{IEEEbiography}[{\includegraphics[width=1in,height=1.25in,clip,keepaspectratio]{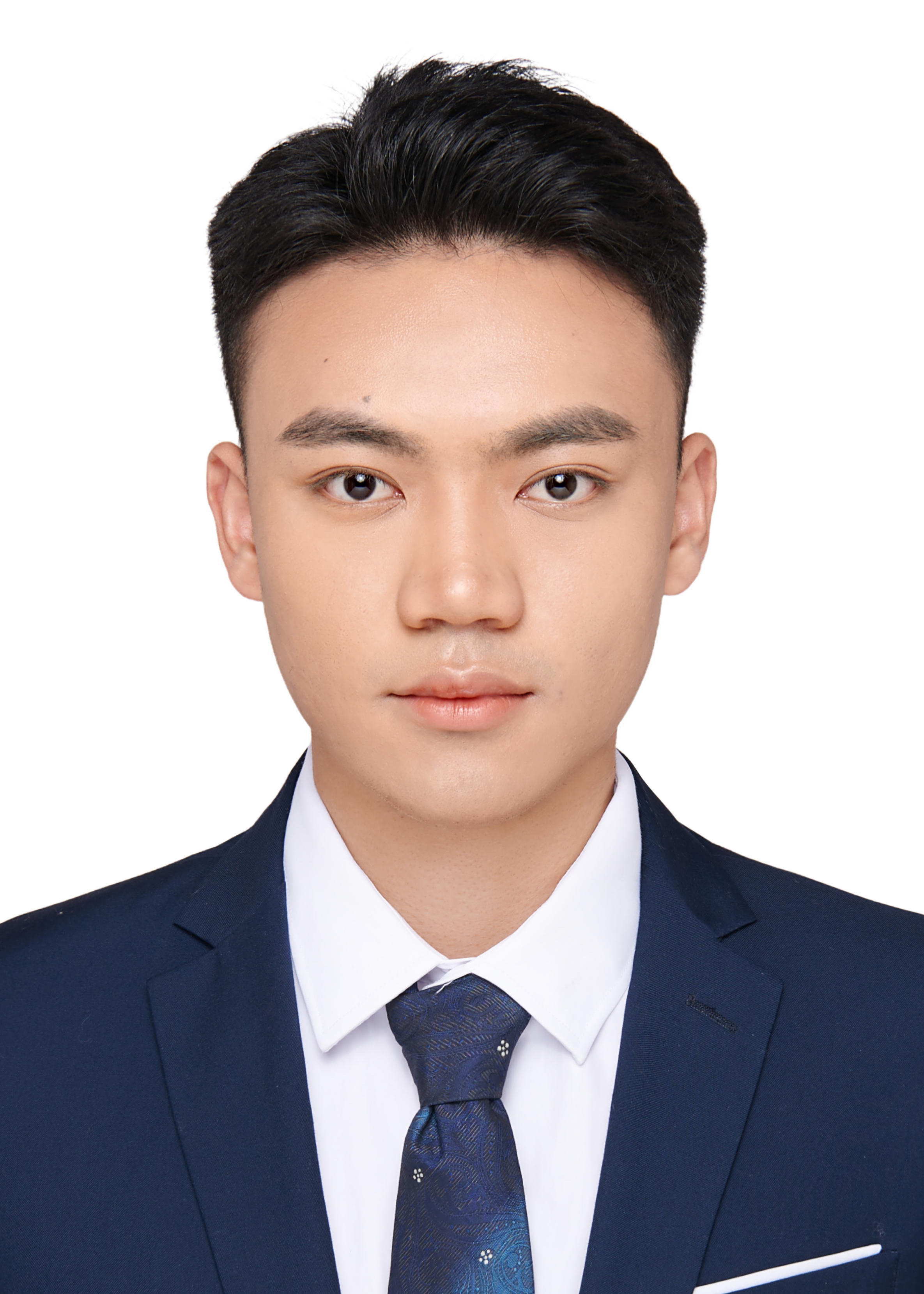}}]{Zihan Wang}
received the B.S. degree from the University of Electronic Science and Technology of China (UESTC) in 2024 and is currently pursuing the Ph.D. degree in Cyber Security at UESTC.
His research interests focus on AI security, with a particular emphasis on the security and safety of LLMs.
\end{IEEEbiography}

\vspace{-10mm} 

\begin{IEEEbiography}[{\includegraphics[width=1in,height=1.25in,clip,keepaspectratio]{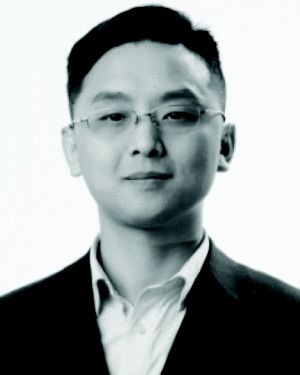}}]{Hongwei Li}
received the PhD degree from the University of Electronic Science and Technology of China, China, in 2008. He is currently a professor with the University of Electronic Science and Technology of China, China. Until October 2012, he was a postdoctoral fellow with the Department of Electrical and Computer Engineering, University of Waterloo for one year. His research interests include network security, applied cryptography, and trusted computing. He was the associate editor for the IEEE Internet of Things Journal, the guest editor of the IEEE Networking, and the Peer to-Peer Networking and Applications. He also was with the technical program committees for many international conferences, including the IEEE INFOCOM, IEEE ICC, IEEE GLOBECOM, IEEE WCNC, IEEE SmartGridComm, BODYNETS, and IEEE DASC. He is the distinguished lecturer of the IEEE Vehicular Technology Society. 
\end{IEEEbiography}
\vspace{-10mm} 

\begin{IEEEbiography}[{\includegraphics[width=1in,height=1.25in,clip,keepaspectratio]{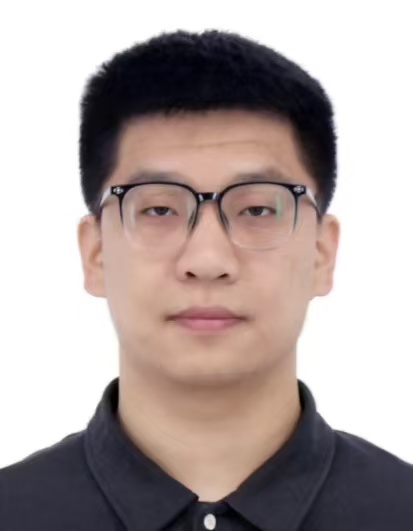}}]{Rui Zhang}
received the B.S. degree in information security in 2020, from the University of Electronic Science and Technology of China (UESTC), where he is currently working toward the Ph.D. degree in cyber security at UESTC. His research interests include AI security and privacy-preserving deep learning. 
\end{IEEEbiography}

\vspace{-10mm} 

\begin{IEEEbiography}[{\includegraphics[width=1in,height=1.25in,clip,keepaspectratio]{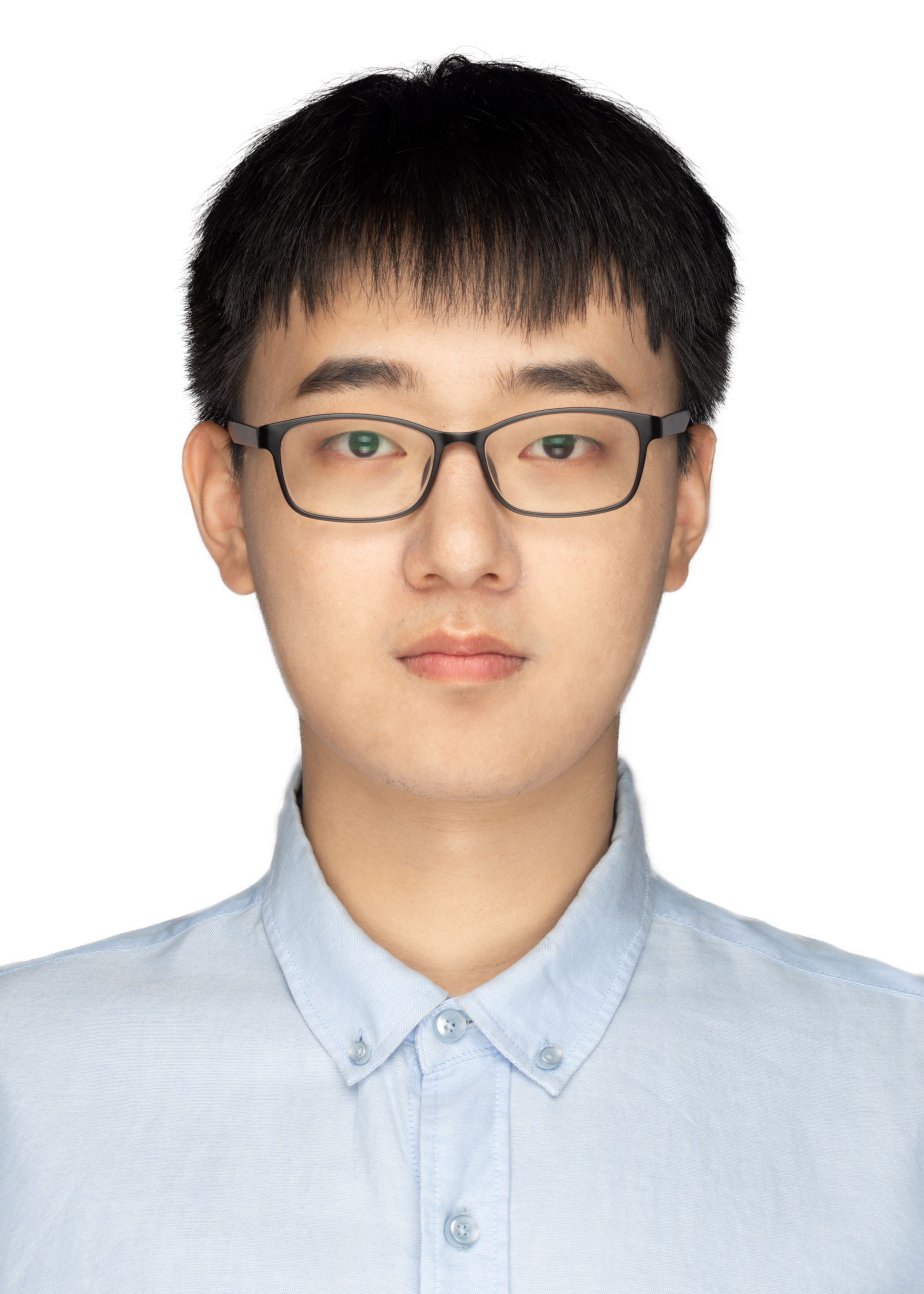}}]{Wenbo Jiang}
is currently a Postdoc at University of Electronic Science and Technology of China (UESTC). He received the Ph.D. degree in cybersecurity from UESTC in 2023 and studied as a visiting Ph.D. student from Jul. 2021 to Jul. 2022 at Nanyang Technological University, Singapore. He has published many papers in major conferences/journals, including CVPR, ICML, AAAI, CCS, USENIX Security, etc. His research interests include trustworthy AI and data security.
\end{IEEEbiography}
\vspace{-10mm}

\begin{IEEEbiography}[{\includegraphics[width=1in,height=1.25in,clip,keepaspectratio]{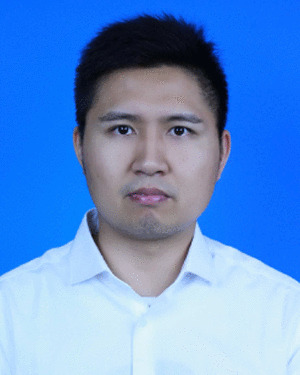}}]{Guowen Xu}
received the PhD degree from the University of Electronic Science and Technology of China, in 2020. He is currently a professor with the School of Computer Science and Engineering (School of Cyber Security), University of Electronic Science and Technology of China. He is the recipient of the Best Paper Award of the 26th IEEE International Conference on Parallel and Distributed Systems (ICPADS 2020), the Best Student Paper Award of the Sichuan Province Computer Federation (SCF 2019), the Student Conference Award of IEEE International Conference on Computer Communications (INFOCOM 2020), and the Distinguished Reviewer of ACM Transactions on the Web. His research interests include applied cryptography and privacy-preserving issues in Deep Learning. He is currently serving as associate editors on IEEE Transactions on Information Forensics and Security (TIFS), IEEE Transactions on Circuits and Systems for Video Technology (TSCVT), IEEE Transactions on Network and Service Management (TNSM) and Pattern Recognition (PR).
\end{IEEEbiography}

\newpage
\appendix

\end{document}